\documentclass[12pt,draftclsnofoot,onecolumn]{IEEEtran}

\IEEEoverridecommandlockouts
\usepackage{indentfirst,flushend}
\usepackage{amssymb,bm,mathrsfs,times,amscd,amsmath,amsthm,amsfonts,bbm}
\usepackage{latexsym}
\usepackage{dsfont,diagbox}
\usepackage{graphicx}
\usepackage{subfigure}
\usepackage{color}
\usepackage{cases}

\newtheorem{remark}{Remark}
\usepackage{multirow}
\usepackage[sort]{cite}
\usepackage{multicol}
\usepackage[nooneline,flushleft]{caption2}
\usepackage{clrscode}
\usepackage{algorithm}
\usepackage{psfrag,stfloats,nicefrac}
\usepackage{algorithmic}
\usepackage{supertabular}
\usepackage{makecell}
\usepackage{epstopdf}
\allowdisplaybreaks[4]

\begin{document}

\title{Two-level Transmission Scheme for Cache-enabled Fog Radio Access Networks}
\author{Shiwen~He,~\IEEEmembership{Member,~IEEE}, Chenhao~Qi,~\IEEEmembership{Senior Member,~IEEE}, Yongming~Huang,~\IEEEmembership{Senior Member,~IEEE}, Qi Hou,\\ and Arumugam Nallanathan,~\IEEEmembership{Fellow,~IEEE}

\thanks{S. He is with the School of Information Science and Engineering, Central South University, Changsha 410083, China. (email: shiwen.he.hn@csu.edu.cn). }

\thanks{C. Qi and Q. Hou are with the School of Information Science and Engineering, Southeast University, Nanjing 210096, China. (email: \{qch,220150860\}@seu.edu.cn). }

\thanks{Y. Huang is with the National Mobile Communications Research Laboratory, Southeast University, Nanjing 210096, China. (email: huangym@seu.edu.cn). }

\thanks{A. Nallanathan is with the School of Electronic Engineering and Computer Science, Queen Mary University of London, UK (email: a.nallanathan@qmul.ac.uk).}
}

\maketitle
\vspace{-.6 in}

\begin{abstract}
In this paper, we investigate the downlink transmission for cache-enabled fog radio access networks aiming at maximizing the delivery rate under the constraints of fronthaul capacity, maximum transmit power, and size of files. To reduce the delivery latency and the burden on fronthaul links and make full use of the local cache and baseband signal processing capabilities of enhanced remote radio heads (eRRHs), a two-level transmission scheme including cache-level and network-level transmission is proposed. In cache-level transmission, only requested files cached at the local cache are transmitted to the corresponding users. The duration of cache-level transmission is the delay caused by the transfer between the baseband unit (BBU) and eRRHs as well as the signal processing at the BBU. The remaining requested files are jointly transmitted to the corresponding users at network-level transmission. For cache-level transmission, a centralized optimization algorithm is firstly presented and then a decentralized optimization algorithm is provided to avoid the exchange of signaling among eRRHs. Meanwhile, another centralized optimization algorithm is presented to tackle the optimization problem for network-level transmission. All presented algorithms are proved to converge to the Karush-Kuhn-Tucker (KKT) solutions of the problems. Numerical results are provided to validate the effectiveness of the proposed transmission scheme as well as evaluating the system performance.
\end{abstract}


\section*{\sc \uppercase\expandafter{\romannumeral1}. Introduction}

To cope with the continuously increasing number of wireless devices, to satisfy the demand of high data rates, and to meet the stringent quality-of-service (QoS) requirement of the emerging wireless services, various advanced communication technologies have been proposed in recent years~\cite{JSACSharfi2017}. For example, cooperative communication~\cite{CMagLee2012}, massive multiple-input multiple-output (MIMO)~\cite{MagBjo2016}, and network densification~\cite{JSACWang2017} are key technologies to achieve high capacity, high data rate, and guarantee network coverage in the fifth generation (5G) mobile cellular systems. Especially for network densification, cloud radio access network (C-RAN) is considered as an emerging network architecture that shows significant promises in suppressing interference and coordinately allocating resource for wireless networks~\cite{TCOMLou2014,WirelssPeng2015,MagPark2014}. In C-RAN, the remote radio heads (RRHs) are connected to a central processor, i.e., baseband signal processing unit (BBU), through fronthaul links~\cite{SurChec2015,NetwSim2016,CMagKu2017}. However, performing completely joint processing requires not only signaling overhead but also payload data sharing among all coordinated RRHs, resulting in tremendous burden on fronthaul links~\cite{TSPPark2013,TVTKang2016,TSPZhou2016,JSACDai2016,SPLPark2016}. Thus, the design of transmission scheme needs to take the limitation of the capacity of fronthaul links into account. In~\cite{TSPPark2013,TVTKang2016,TSPZhou2016}, the maximization problem of weighted sum rate was investigated under the constraint of fronthaul capacity. In~\cite{JSACDai2016}, the data-sharing strategy and compression strategy were studied regarding the energy efficiency of the C-RANs. In~\cite{SPLPark2016}, joint design of fronthaul and radio access links for C-RANs was investigated. Note that a common assumption in aforementioned literature is that the coordinated RRHs do not have the ability to cache data files.

One of the most challenging requirements in 5G is the provision of connections with low end-to-end latency. In most cases, the latency should be smaller than 100 ms, while sometimes it is smaller than 10 ms~\cite{5GWhitePaper}. The low latency is demanded by some applications such as augmented reality and vehicular communications~\cite{CSTHayat2016}. One solution to reduce the end-to-end latency as well as the fronthaul bottleneck in C-RANs is using distributed caches at the edge of the mobile cellular network, referred to as fog radio access networks (F-RANs)~\cite{CMagTran2017,TVTHou2016,CMagWang2014}. The resulting RRHs are referred to as enhanced RRHs (eRRHs). The eRRHs can pre-fetch the most frequently requested files to the eRRHs' local caches during off-peak traffic periods so that the fronthaul overhead can be reduced during the peak traffic periods. In this way, lower latency and higher spectral efficiency can be achieved~\cite{CMagWang2014}. In~\cite{CLZheng2017}, the optimization of hybrid cache placement of data files for coordinated relaying networks was studied. Then the cache placement in F-RANs was investigated by fully considering the flexibility of physical-layer transmission and diverse content preferences of different users~\cite{TWCLiu2017}. In~\cite{TWCPark2016}, joint design of cloud and edge processing for the downlink of F-RANs was investigated for maximizing the minimum user delivery rate. In~\cite{TWCTao2016}, the problem of minimizing the system cost that accounts for fronthaul overhead while satisfying the target signal-to-interference-plus-noise ratio (SINR) was considered. More recently, the authors of~\cite{TWCXiang2018} proposed to use wireless caching to enhance the physical layer security of video streaming in cellular networks with limited backhaul capacity.

In general, the sum rate maximization or power minimization problem subject to QoS and transmit power constraints needs to be centrally solved for interference channels~\cite{TWCPark2016,TWCTao2016,TWCXiang2018}. Meanwhile, some decentralized optimization algorithms were also developed to reduce the burden of information collection~\cite{TSPShi2011,TSPWeeraddana2013,TSPHuang2011,TSPTervo2018,TVTHuang2014}. In~\cite{TSPShi2011}, a distributed weighted sum rate maximization (WSRMax) method was proposed via exploiting the relation between the user rate and the minimum mean square error (MMSE) for interfering broadcast channel. At each iteration all mobile terminals need to first estimate the covariance matrices of their received signals, and then compute and feed back some parameters to the eRRHs. To avoid the estimation of the covariance matrices at each mobile terminal, the authors of~\cite{TSPWeeraddana2013} proposed a distributed approach to maximize the WSRMax problem with limited signaling exchange among eRRHs. In~\cite{TSPTervo2018}, both centralized and distributed beamforming algorithms were proposed to solve power minimization and SINR balancing problem. In~\cite{TSPHuang2011}, the max-min fairness problem under the total power constraint was tackled from the perspective of transmit power minimization, where a distributed hierarchical iterative algorithm was proposed based on the uplink-downlink duality. In~\cite{TVTHuang2014}, a decentralized optimization method was proposed by exploiting interference temperature~\cite{TSPZhang2010} to maximize the system energy efficiency. However, the decentralized optimization algorithms in the aforementioned literature need signaling exchange in each iteration during the optimization process, which not only requires strict synchronization but also generates some backhaul or fronthaul burden among coordinated eRRHs. The signaling exchange per iteration between coordinated eRRHs is nevertheless not desirable for low latency communications.

The joint transmission of cached and uncached requested contents in~\cite{TWCPark2016,TWCTao2016,TWCXiang2018} only used the caching to reduce the burden on fronthaul links without making full use of the caching to reduce the delivery latency. However, one of the main motivation of introducing edge caching is to reduce the delivery latency for satisfying the demands of the latency-sensitive traffic for future wireless communication system~\cite{TITSeng2017}. To the best of the authors' knowledge, how to fully exploit the capabilities, i.e., caching and baseband signal processing, of eRRHs to reduce the delivery latency and the overhead of fronthaul links is still an enormous challenge for wireless transmission in F-RANs. In this paper, we investigate the downlink transmission mechanism for cache-enabled F-RANs. We divide the transmission of the requested files into two-level, i.e., cache-level transmission and network-level transmission. In cache-level transmission, only requested files cached at the local cache are transmitted to the corresponding users. The duration of cache-level transmission is the delay caused by the transfer between the BBU and eRRHs as well as by the signal processing at the BBU. The remaining requested files are jointly transmitted to the corresponding users at network-level transmission. Our objective is to maximize the delivery rate under the constraints of fronthaul capacity, maximum transmit power of each eRRH, and size of cached data files. The corresponding optimization algorithms are presented to address the problems of interest. Specifically, the contribution of this work is summarized as follows:
\begin{itemize}
\item A two-level transmission scheme including cache-level and network-level transmission is proposed to reduce the delivery latency and the burden on fronthaul links by making full use of the local cache and the signal processing capabilities of eRRHs.
\item For cache-level transmission, coordinated beamforming is adopted to improve the spectral efficiency. A centralized optimization algorithm is firstly designed to maximize the sum of network delivery rates. To avoid the demand of centralized processing and the exchange of signaling among eRRHs, a decentralized algorithm is then presented to maximize the sum of network delivery rates by introducing a new concept of signal-to-interference-leakage-noise ratio (SILNR) for cache-level transmission.
\item For network-level transmission, joint transmission is adopted to maximize the system performance such that the inter-user interference can be effectively controlled. At the same time, an optimization algorithm is developed using convex approximation methods.
\item All presented algorithms are proved to converge to the Karush-Kuhn-Tucker (KKT) solutions of the problems. The computational complexity of all presented algorithms is also analyzed. Numerical results are further provided to corroborate the effectiveness of proposed algorithms and unveil that how to cache files and how many files to cache are key problems for cache-enabled F-RANs.
\end{itemize}

The remainder of this paper is organized as follows. The system model is described in Section~\uppercase\expandafter{\romannumeral2}. Section~\uppercase\expandafter{\romannumeral3} formulates the optimization problem of the proposed two-level transmission scheme. The design of optimization algorithm for cache-level transmission is investigated in Section~\uppercase\expandafter{\romannumeral4}. In Section~\uppercase\expandafter{\romannumeral5}, an optimization algorithm for network-level transmission is developed. Numerical results are provided in Section~\uppercase\expandafter{\romannumeral6}. Finally, Section~\uppercase\expandafter{\romannumeral7} concludes the paper.

\textbf{Notations}: Bold lowercase and uppercase symbols represent column vectors and matrices, respectively. The superscripts $\left(\cdot\right)^{T}$ and $\left(\cdot\right)^{H}$ represent the matrix transpose and conjugate transpose, respectively. For a set $\mathcal{A}$, $\left|\mathcal{A}\right|$ denotes the cardinality of $\mathcal{A}$. For a complex-valued number $x$ and a matrix $\mathbf{A}$, $\left|x\right|$ and $\left|\mathbf{A}\right|$ denote the absolute value of $x$ and the determinant of $\mathbf{A}$, respectively. $\mathrm{tr}\left(\cdot\right)$ and $\|\cdot\|$ denote the trace and the Euclidean norm, respectively. $\mathbf{0}_{N\times N}$ and $\mathbf{I}_{N\times N}$ denote the $N\times N$ zero matrix and $N\times N$ identity matrix, respectively. The circularly symmetric complex Gaussian distribution with mean $\mathbf{\mu}$ and covariance matrix $\mathbf{R}$ is denotes by $\mathcal{CN}\left(\mathbf{\mu}, \mathbf{R}\right)$. $\mathrm{Diag}(\mathbf{a})$ denotes a diagonal matrix with the main diagonal given by $\mathbf{a}$.

\section*{\sc \uppercase\expandafter{\romannumeral2}. System Model}
\begin{figure}[h]
\renewcommand{\captionfont}{\footnotesize}
\renewcommand*\captionlabeldelim{.}
\centering
\captionstyle{flushleft}
\onelinecaptionstrue
\includegraphics[width=0.8\columnwidth,keepaspectratio]{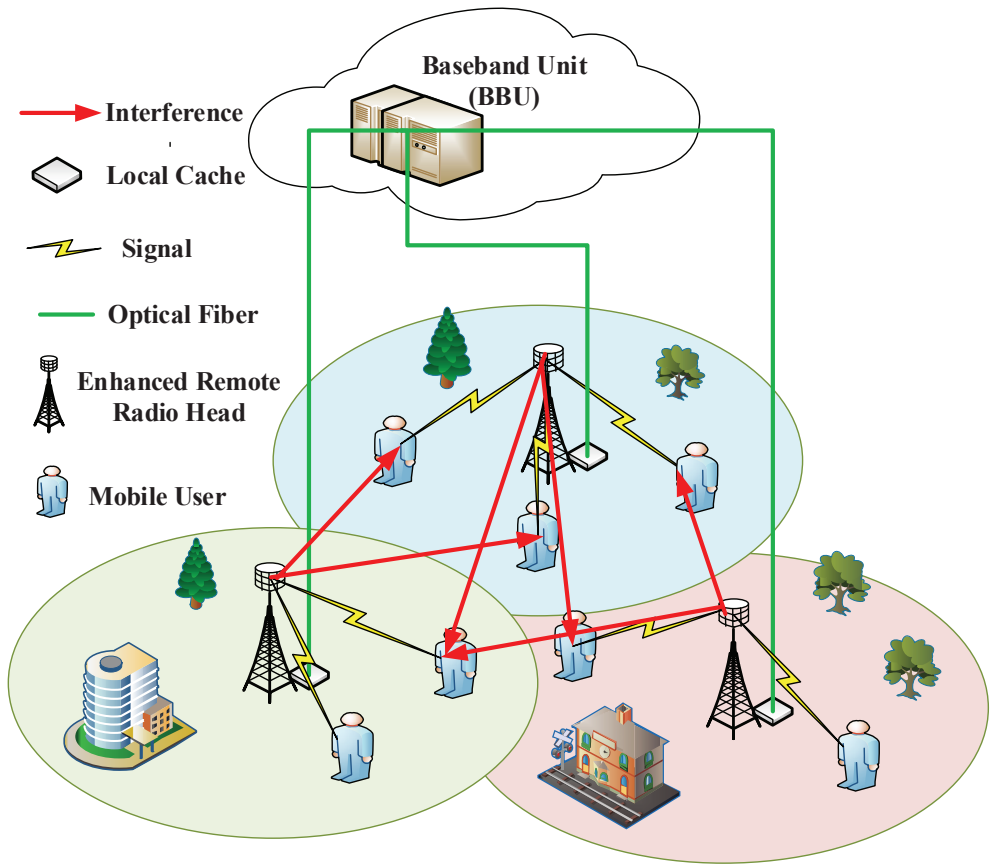}\\
\caption{Illustration of the system model.}
\label{EquivalentSystemModel}
\end{figure}

As illustrated in Fig.~\ref{EquivalentSystemModel}, the considered cache-enabled F-RANs include a BBU connected to $K_{\mathrm{R}}$ eRRHs serving totally $K_{\mathrm{U}}$ single-antenna users. The index sets of the eRRHs and the users are denoted as $\mathcal{K}_{\mathrm{R}} \triangleq \left\{1,\cdots,K_{\mathrm{R}}\right\}$ and $\mathcal{K}_{\mathrm{U}} \triangleq \left\{1,\cdots,K_{\mathrm{U}}\right\}$, respectively. The eRRH $i$ equipped with $N_{\mathrm{t},i}$ antennas is connected to the BBU through an error-free fronthaul link with the capacity of $C_{i}$ bits per symbol. The eRRH $i$ is also equipped with a cache, which can store $nB_{i}>0$ bits, $\forall i\in\mathcal{K}_{\mathrm{R}}$, where $n$ is the number of symbols transmitted in each downlink transmission interval. The BBU in the ``cloud'' can perform signal processing globally, while the eRRHs at the edge of the ``cloud" can only perform signal processing locally.

Assume that each user $k\in\mathcal{K}_{\mathrm{U}}$ requests contents or files from library $\mathcal{F}=\left\{1,\cdots,F\right\}$ stored in the BBU. Without loss of generality, we assume that all files in library $\mathcal{F}$ are of the same size of $nS$ bits. For simplicity, in this paper, we further assume that the files cannot be partitioned and different eRRH caches different files. The eRRH $i\in\mathcal{K_{\mathrm{R}}}$ pre-stores $\lfloor B_{i}/S \rfloor$ files in its cache based on the user behavior, the delay-sensitive requirement, the long-term information of the popularity distribution, the fronthaul capacity, and etc~\cite{TWCPark2016,TWCTao2016,CLZheng2017,TWCLiu2017}. The state of cached file $f$, $f\in\mathcal{F}$, can be modeled by defining binary variables $c_{f,i}$, $f\in\mathcal{F}$, $i\in\mathcal{K}_{\mathrm{R}}$, as

\begin{equation}\label{DelayAware02}
c_{f,i}=
\begin{cases}
1, ~~\text{if}~f~\text{is cached by eRRH}~i,\\
0 ,~~\text{otherwise} .
\end{cases}
\end{equation}
At the same time, $c_{f,i}$ satisfies the conditions $\sum\limits_{i\in\mathcal{K}_{\mathrm{R}}}c_{f,i}=0$ or $\sum\limits_{i\in\mathcal{K}_{\mathrm{R}}}c_{f,i}=1$, indicating that different eRRH caches different files. For simplicity, in this work, we assume that the state information $c_{f,i}$, $f\in\mathcal{F}$, $i\in\mathcal{K}_{\mathrm{R}}$, is predetermined. Thus, if file $f_{k}$ requested by user $k$ is cached in the  local cache of its serving eRRH, it can be retrieved directly from the eRRH without downloading from the BBU\footnote{Each eRRH can retrieve the information of requested files by learning-based methods and is out of the scope of this work. The developed method in this work can be easily extended to the case where each eRRH caches multiple different files and the case where the file can be partitioned into multiple subfiles.}. Otherwise, the information of file $f_{k}$ needs to be transferred to its serving eRRH from the BBU via the fronthaul link.

\setcounter{equation}{4}
\begin{figure*}
\hrulefill
\begin{equation}\label{DelayAware05}
\gamma_{k,i}^{\mathrm{c}}=\frac{\left|c_{f_{k},i}\mathbf{h}_{k,i}^{H}\mathbf{g}_{k,i}\right|^{2}}
{\sum\limits_{k'\in\mathcal{K}_{i}\setminus\left\{k\right\}}\left|c_{f_{k'},i}\mathbf{h}_{k,i}^{H}\mathbf{g}_{k',i}\right|^{2}+
\sum\limits_{j\in\mathcal{K}_{\mathrm{R}}\setminus\left\{i\right\}}\sum\limits_{k'\in\mathcal{K}_{j}}\left|c_{f_{k'},j}\mathbf{h}_{k,j}^{H}\mathbf{g}_{k',j}\right|^{2}+\sigma^{2}}.
\end{equation}
\hrulefill
\end{figure*}
\section*{\sc \uppercase\expandafter{\romannumeral3}. Problem Formulations}

In order to make full use of the functionalities of cache as well as the signal processing capability of each eRRH, in this paper, the data transmission is divided into two-level transmission. The first level transmission is cache-level transmission where all eRRHs coordinately transmit cached files to scheduled users via coordinated beamforming. The duration of cache-level transmission is the delay caused by the signal processing at the BBU and the information exchange between the eRRH and the BBU. The second level transmission is network-level transmission where a quantized version of the precoded signals of the untransmitted requested files is sent to the eRRHs via fronthaul links and then all eRRHs jointly transmit it to $K_{\mathrm{U}}$ users via joint transmission.
\subsection*{A. Cache-level Transmission}
To reduce the latency incurred by the propagation of the fronthaul link and the signal processing at the BBU, eRRH $i$ firstly transmits data to users, denoted as $\mathcal{K}_{i}\subseteq\mathcal{K}_{\mathrm{U}}$, whose requested files are stored in the local cache in cache-level transmission. Furthermore, $\mathcal{K}_{i}\cap \mathcal{K}_{j}\neq\emptyset$, $i\neq j\in\mathcal{K}_{\mathrm{R}}$, and $\cup_{i\in\mathcal{K}_{\mathrm{R}}}\mathcal{K}_{i}\subseteq \mathcal{K}_{\mathrm{U}}$. As a numerical example, user set $\mathcal{K}_{i}$ can be given by
\setcounter{equation}{1}
\begin{equation}\label{DelayAware03}
\mathcal{K}_{i}=\left\{k: \left| \left\{k': c_{f_{k'},i} \left\|\mathbf{h}_{k',i}\right\|> c_{f_{k},i} \left\|\mathbf{h}_{k,i}\right\| \right\}\right| < N_{\mathrm{s}}\atop~\text{and}~c_{f_{k},i}==1\right\}
\end{equation}
where $N_{\mathrm{s}}$ is a scalable parameter with $N_{\mathrm{s}}\leq N_{\mathrm{t},i}$ and $f_{k}$ is the requested file index of user $k$. Here we assume that each user requests a single file. For cache-level transmission, the signal transmitted by the eRRH $i$ can be expressed as
\begin{equation}\label{DelayAware04}
\mathbf{x}_{i}^{\mathrm{c}}=\sum\limits_{k\in\mathcal{K}_{i}}c_{f_{k},i}\mathbf{g}_{k,i}s_{f_{k}}.
\end{equation}
where $\mathbf{g}_{k,i}\in\mathbb{C}^{N_{\mathrm{t},i}\times 1}$ is the beamforming vector and $\mathbf{g}_{k,i}=\mathbf{0}$ for $k\notin\mathcal{K}_{i}$. The signal received at user $k$, $\forall k\in\mathcal{K}_{\mathrm{U}}$, can be expressed as:
\begin{equation}\label{DelayAware01}
y_{k}^{\mathrm{c}}=\sum\limits_{i\in\mathcal{K}_{\mathrm{R}}}\mathbf{h}_{k,i}^{H}\mathbf{x}_{i}^{\mathrm{c}}+n_{k},
\end{equation}
where $\mathbf{h}_{k,i}\in\mathbb{C}^{N_{\mathrm{t},i}\times 1}$ denotes the channel coefficients between user $k$ and eRRH $i$ and $n_{k}\sim\mathcal{CN}\left(0,\sigma^{2}\right)$ is the additive white Gaussian noise (AWGN). Note that in~\eqref{DelayAware01}, we assume that baseband signal $s_{f_{k}}$ and noise $n_{k}$ are independent, $\forall i\in\mathcal{K}_{{\mathrm{R}}}, k\in\mathcal{K}_{{\mathrm{U}}}$. Thus, the achievable rate of user $k\in\mathcal{K}_{\mathrm{U}}$ served by eRRH $i$ is calculated as $\mathrm{R}_{k,i}^{\mathrm{c}}=\log_{2}\left(1+\gamma_{k,i}^{\mathrm{c}}\right)$, where $\gamma_{k,i}^{\mathrm{c}}$ is the signal-to-interference-plus-noise ratio (SINR) that is calculated as~\eqref{DelayAware05}, at the top of the next page.

The goal is to maximize the sum of delivery rate by optimizing the transmit beamforming vectors at each level transmission. Thus, for cache-level transmission, the optimization of $\mathbf{g}_{k,i}$, $k\in\mathcal{K}_{i}$, $i\in\mathcal{K}_{\mathrm{R}}$, can be realized by solving the following problem\footnote{The idea of this work can be extended to the scenario where multiple users request the same file via multi-cast transmission as well as the scenario where the user equipped with multiple antennas requests multiple files via spatial multiplexing. What we need to do is to adjust the formulation of problems~\eqref{DelayAware08} and~\eqref{DelayAware09} according to the parameters of the corresponding scenario. However, it is left as our future work.}:
\setcounter{equation}{5}
\begin{subequations}\label{DelayAware08}
\begin{align}
&\max~\tau\sum\limits_{i\in\mathcal{K}_{\mathrm{R}}}\sum\limits_{k\in\mathcal{K}_{i}}\mathrm{r}_{k,i}^{\mathrm{c}},\label{DelayAware08a}\\
s.t. ~&\mathrm{r}_{k,i}^{\mathrm{c}}\leq\mathrm{R}_{k,i}^{\mathrm{c}}, \forall k\in\mathcal{K}_{i}, \forall i\in\mathcal{K}_{\mathrm{R}},\label{DelayAware08b}\\
&\tau\mathrm{r}_{k,i}^{\mathrm{c}}\leq S, \forall k\in\mathcal{K}_{i}, \forall i\in\mathcal{K}_{\mathrm{R}},\label{DelayAware08c}\\
&\sum\limits_{k\in\mathcal{K}_{i}}\left\|c_{f_{k},i}\mathbf{g}_{k,i}\right\|^{2}\leq P_{i}, \forall i\in\mathcal{K}_{\mathrm{R}},\label{DelayAware08d}
\end{align}
\end{subequations}
where the variable is $\left\{\mathbf{g}_{k,i},\mathrm{r}_{k,i}^{\mathrm{c}}\right\}_{k\in\mathcal{K}_{i}, i\in\mathcal{K}_{\mathrm{R}}}$. $\mathrm{r}_{k,i}^{\mathrm{c}}$ denotes the delivery rate of user $k$ served by eRRH $i$, $\forall k\in\mathcal{K}_{i}$, $\forall i\in\mathcal{K}_{\mathrm{R}}$, for cache-level transmission. In problem~\eqref{DelayAware08}, constraint~\eqref{DelayAware08b} shows that the delivery rate $\mathrm{r}_{k,i}^{\mathrm{c}}$ is bounded by the achievable data rate $\mathrm{R}_{k,i}^{\mathrm{c}}$. Constraint~\eqref{DelayAware08c} denotes that the number of delivery data cannot exceed the cached file size $S$. Constraint~\eqref{DelayAware08d} means the maximum transmit power constraint on the beamforming vectors. Delay $\tau$ is a summation of the signal processing time at the BBU and the information exchange time between eRRH $i$ and the BBU\footnote{For simplicity, we assume that the delay $\tau$ is the same for all eRRHs. For full caching case, i.e., all requested files are stored the local cache, the value of the delay $\tau$ is set be unit in optimization problem~\eqref{DelayAware08} and network-level transmission is not need.}. From the description of problem~\eqref{DelayAware08}, one can see that the duration of cache-level transmission is also the delay $\tau$.

\subsection*{B. Network-level Transmission}
For network-level transmission, all eRRHs coordinately transmit data to all scheduled users by using joint transmission~\cite{CMagLee2012}. Although joint transmission can effectively control the intra- and inter-eRRH interference and improve the quality of service for each user, the BBU needs to collect all channel state information (CSI) and allocates the processed data files to each eRRH  resulting in a considerable large amount of overhead on the fronthaul links. For network-level transmission, the signal transmitted by eRRH $i$ is expressed as\footnote{Note that point-to-point compression algorithm is adopted to quantize the precoded signals for each eRRH during the network-level transmission. As pointed out in~\cite{MagPark2014}, we can adopt some advanced compression algorithm to quantize the precoded signals such that the system performance can be further improved. Therefore, we would like to study the performance of the proposed two-level transmission scheme with advanced compression algorithm in our future work.}:
\begin{equation}\label{DelayAware06}
\mathbf{x}_{i}^{\mathrm{n}}=\mathbf{x}_{i}+\mathbf{q}_{i}=
\sum\limits_{k\in\mathcal{K}_{\mathrm{U}}}\mathbf{u}_{k,i}s_{f_{k}}+\mathbf{q}_{i},
\end{equation}
where $\mathbf{u}_{k,i}\in\mathds{C}^{N_{\mathrm{t},i}\times 1}$ is the beamforming vector for file $f_{k}$ used at eRRH $i$. The quantization noise $\mathbf{q}_{i}\in\mathds{C}^{N_{\mathrm{t},i}\times 1}$ is assumed to be independent of the transmitting signal and distributed as $\mathbf{q}_{i}\sim\mathcal{CN}\left(\mathbf{0},\mathbf{\Omega}_{i}\right)$. The signals $\mathbf{x}_{i}$ and $\mathbf{x}_{j}$ for different eRRHs $i \neq j$ are quantified independently so that the quantization noise signals $\mathbf{q}_{i}$ and $\mathbf{q}_{j}$ are independent~\cite{TWCPark2016}. For ease of presentation, let $\mathbf{\Omega}$ be the covariance matrix of quantization noise, i.e.,  $\mathbf{\Omega}=\mathrm{Diag}\left(\mathbf{\Omega}_{1},\cdots,\mathbf{\Omega}_{K_{\mathrm{R}}}\right)$. For network-level transmission, the signal received at user $k$, $\forall k\in\mathcal{K}_{\mathrm{U}}$, can be expressed as:
\begin{equation}\label{DelayAware010}
y_{k}^{\mathrm{n}}=\sum\limits_{i\in\mathcal{K}_{\mathrm{R}}}\mathbf{h}_{k}^{H}\mathbf{x}+n_{k},
\end{equation}
where $\mathbf{h}_{k}=\left[\mathbf{h}_{k,1}^{H},\cdots,\mathbf{h}_{k,K_{\mathrm{R}}}^{H}\right]^{H}$ and $\mathbf{x}=\left[\left(\mathbf{x}_{i}^{\mathrm{n}}\right)^{H},\cdots,\left(\mathbf{x}_{K_{\mathrm{R}}}^{\mathrm{n}}\right)^{H}\right]^{H}$. For the joint transmission, we implicitly assume that the stringent synchronization of timing and frequency among eRRHs has been finished. Similarly, the achievable rate of user $k\in\mathcal{K}_{\mathrm{U}}$ is calculated as $\mathrm{R}_{k}^{\mathrm{n}}=\log_{2}\left(1+\gamma_{k}^{\mathrm{n}}\right)$, where SINR $\gamma_{k}^{\mathrm{n}}$ is given by:
\begin{equation}\label{DelayAware07}
\gamma_{k}^{\mathrm{n}}=\frac{\left|\mathbf{h}_{k}^{H}\mathbf{u}_{k}\right|^{2}}
{\sum\limits_{k'\in\mathcal{K}_{\mathrm{U}}, k'\neq k}\left|\mathbf{h}_{k}^{H}\mathbf{u}_{k'}\right|^{2}+\mathbf{h}_{k}^{H}\mathbf{\Omega}\mathbf{h}_{k}+\sigma^{2}},
\end{equation}
where $\mathbf{u}_{k}=\left[\mathbf{u}_{k,1}^{\mathrm{H}},\cdots,\mathbf{u}_{k,K_{\mathrm{R}}}^{\mathrm{H}}\right]^{\mathrm{H}}$.

Similarly, for network-level transmission, the transmit beamforming vectors $\mathbf{u}_{k}$, $k\in\mathcal{K}_{\mathrm{U}}$, are optimized centrally at the BBU by addressing the following problem:
\begin{subequations}\label{DelayAware09}
\begin{align}
&\max~\sum\limits_{k\in\mathcal{K}_{\mathrm{U}}}\mathrm{r}_{k}^{\mathrm{n}},\label{DelayAware09a}\\
s.t. ~&\mathrm{r}_{k}^{\mathrm{n}}\leq\mathrm{R}_{k}^{\mathrm{n}},  \forall k\in\mathcal{K}_{U},\label{DelayAware09b}\\
&\sum\limits_{k\in\mathcal{K}_{\mathrm{U}}}\left\|\mathbf{B}_{i}\mathbf{u}_{k}\right\|^{2}+\mathrm{tr}\left(\mathbf{\Omega}_{i}\right)\leq P_{i}, \forall i\in\mathcal{K}_{R},\label{DelayAware09c}\\
&\mathrm{r}_{k}^{\mathrm{n}}\leq \max\left(S-\tau\overline{\mathrm{r}}_{k}^{\mathrm{c}},0\right), \forall k\in\mathcal{K}_{U},\label{DelayAware09d}\\
&\log_{2}\left(\left|\overline{\mathbf{A}}_{i}\right|\right)-\log_{2}\left(\left|\mathbf{\Omega}_{i}\right|\right)\leqslant C_{i}, \forall i\in\mathcal{K}_{R},\label{DelayAware09e}
\end{align}
\end{subequations}
where the variables are $\left\{\mathbf{u}_{k},\mathrm{r}_{k}^{\mathrm{n}}, \mathbf{\Omega}_{i}\right\}_{k\in\mathcal{K}_{\mathrm{U}},  i\in\mathcal{K}_{\mathrm{R}}}$. Also note that $\mathrm{r}_{k}^{\mathrm{n}}$ denotes the delivery rate of user $k$ during network-level transmission, $\forall k\in\mathcal{K}_{\mathrm{U}}$, and  $\overline{\mathbf{A}}_{i}=\mathbf{B}_{i}\sum\limits_{k\in\mathcal{K}_{\mathrm{U}}}\left(1-c_{f_{k},i}\right)\mathbf{u}_{k}
\mathbf{u}_{k}^{\mathrm{H}}\mathbf{B}_{i}^{\mathrm{H}}+\mathbf{\Omega}_{i}$ with matrix $\mathbf{B}_{i}\in\mathbb{C}^{N_{\mathrm{t},i}\times N_{\mathrm{t}}}$ being defined as
$$\mathbf{B}_{i}=\left[\mathbf{0}_{N_{\mathrm{t},i}\times N_{\mathbf{B}_{i}}^{1}},\mathbf{I}_{N_{\mathrm{t},i}\times N_{\mathrm{t},i}},\mathbf{0}_{N_{\mathrm{t},i}\times N_{\mathbf{B}_{i}}^{2}}\right]$$
with $N_{\mathbf{B}_{i}}^{1}=\sum\limits_{j=1}^{i-1}N_{\mathrm{t},j}$ and $N_{\mathbf{B}_{i}}^{2}=N_{\mathrm{tt}}-\sum\limits_{j=1}^{i}N_{\mathrm{t},j}$, where $N_{\mathrm{tt}}=\sum\limits_{i\in\mathcal{K}_{\mathrm{R}}}N_{\mathrm{t},i}$. $\overline{\mathrm{r}}_{k}^{\mathrm{c}}=\max\limits_{i\in\mathcal{K}_{\mathrm{R}}}\mathrm{r}_{k,i}^{\mathrm{c}}$ denotes the maximum transmission rate of user $k$ at cache-level transmission\footnote{We assume that the BBU can obtain the values of $\mathrm{r}_{k,i}^{\mathrm{c}}$, $\forall k\in\mathcal{K}_{\mathrm{U}}$, $\forall i\in\mathcal{K}_{\mathrm{R}}$, according to the cache state information $\left\{c_{f,i}\right\}_{f\in\mathcal{F},i\in\mathcal{K}_{\mathrm{R}}}$ and the transmission scheme adopted by each eRRH, i.e., \eqref{DelayAware08}.}. In problem~\eqref{DelayAware09}, \eqref{DelayAware09d} constrains the size of network-level transmission file not to exceed the remaining file size $\max\left(S-\tau\overline{\mathrm{r}}_{k}^{\mathrm{c}},0\right)$\footnote{Note that different from~\cite{TSPPark2013,TVTKang2016,TSPZhou2016,JSACDai2016},\cite{TWCPark2016}, and~\cite{TWCTao2016}, the size of transmitted files in cache-level transmission is included in the constraints~\eqref{DelayAware09d}. This provides a chance to increase transmit rate of untransmitted files, i.e., improve the system delivery rate.}. Constraint~\eqref{DelayAware09e} is due to the fronthaul capacity constraints ensuring the signal $\mathbf{x}_{i}^{\mathrm{n}}$ can be reliably recovered by eRRH $i$~\cite[Ch~3]{BookGamal2011}. Note that when the delay $\tau$ is zero, problem~\eqref{DelayAware09} becomes the optimization problem of precoding matrices for traditional cloud-based coordinated joint transmission system under the constraints of the fronthaul capacity, the size of  data files, and the maximum transmit power of eRRHs.

From the aforementioned analysis, one can see that the minimum delivery content of our proposed two-level transmission scheme is the sum of the minimum delivery rate achieved by cache-level transmission and that of network-level transmission, i.e., $\mathcal{T}_{\mathrm{p}}=\tau\sum\limits_{i\in\mathcal{K}_{\mathrm{R}}}\sum\limits_{k\in\mathcal{K}_{i}}\mathrm{r}_{k,i}^{\mathrm{c}}+\sum\limits_{k\in\mathcal{K}_{\mathrm{U}}}\mathrm{r}_{k}^{\mathrm{n}}$. The total minimum delivery content is $\mathcal{T}_{\mathrm{T}}=\sum\limits_{k\in\mathcal{K}_{\mathrm{U}}}\mathrm{r}_{k}^{\mathrm{n}}$ for the traditional cloud-based coordinated joint transmission system.
\begin{remark}
\rm Note that in network-level transmission, all eRRHs transmit quantized precoding signals received from the BBU. However, in~[21], all eRRHs transmit a superposition of precoded signals received from the BBU and locally precoded signals based on the caching files. The coupling of the precoding matrices used at all eRRHs make the optimization problem more difficult such that a central controller is used to optimize the precoding matrices~[21]. Thus, the precoding matrices used at each eRRH need to be transmitted to the corresponding eRRH. In our proposed scheme, we use the caching files to simultaneously reduce the delivery latency and the burden on fronthaul links. Namely, in cache-level transmission, the required files which are cached at eRRHs are first transmitted to the corresponding users during the duration of the delay $\tau$. Meanwhile, the burden on fronthaul links is also reduced under the constraints of~\eqref{DelayAware09d} and~\eqref{DelayAware09e}.
\end{remark}

\section*{\sc \uppercase\expandafter{\romannumeral4}. Design of Optimization Algorithm for Cache-level Transmission}

In this section, we first focus on addressing problem~\eqref{DelayAware08} by designing a centralized optimization method. Then, to avoid the exchange of signaling between coordinated eRRHs, a novel concept of signal-to-interference-leakage-noise ratio (SILNR) is introduced to maximize the intra-eRRH delivery rate while suppressing inter-eRRH interference. Finally, we design a decentralized optimization algorithm to obtain the coordinated beamforming vectors for cache-level transmission.

\subsection*{A. Centralized Optimization for Cache-level Transmission}

For coordinated interference networks, the existing of intra-and inter-eRRH interference leads the user rate $\mathrm{R}_{k,i}^{\mathrm{c}}$ to be a nonconcave function. In other words, it is  generally difficult to find the global optimal solution of~\eqref{DelayAware08}. Therefore, this requires the relaxations of the optimization conditions in order to provide reasonable design for practical implementations. The first thing of addressing problem~\eqref{DelayAware08} is to transfer it into a solvable form by using some mathematical methods. In the sequel, we introduce some auxiliary variables $\epsilon_{k,i}$, $\forall k\in\widetilde{\mathcal{K}}_{i}$, $\forall i\in\mathcal{K}_{\mathrm{R}}$, which denote the interference leakage limitation~\cite{JSAC2014He}. Thus, problem~\eqref{DelayAware08} is equivalently reformulated as:
\begin{subequations}\label{DelayAware10}
\begin{align}
&\max~\sum\limits_{i\in\mathcal{K}_{\mathrm{R}}}\sum\limits_{k\in\mathcal{K}_{i}}\mathrm{r}_{k,i}^{\mathrm{c}},\label{DelayAware10a}\\
s.t. ~&\mathrm{r}_{k,i}^{\mathrm{c}}\leq\widehat{\mathrm{R}}_{k,i}^{\mathrm{c}}, \forall k\in\mathcal{K}_{i}, \forall i\in\mathcal{K}_{\mathrm{R}},\label{DelayAware10b}\\
&\sum\limits_{k\in\mathcal{K}_{i}}\left|c_{f_{k},i}\mathbf{h}_{k',i}^{H}\mathbf{g}_{k,i}\right|^{2}\leqslant\epsilon_{k',i}, \forall k'\in\widetilde{\mathcal{K}}_{i}, \forall i\in\mathcal{K}_{\mathrm{R}},\label{DelayAware10c}\\
&\tau\mathrm{r}_{k,i}^{\mathrm{c}}\leq S, \forall k\in\mathcal{K}_{i}, \forall i\in\mathcal{K}_{\mathrm{R}},\label{DelayAware10d}\\
&\sum\limits_{k\in\mathcal{K}_{i}}\left\|c_{f_{k},i}\mathbf{g}_{k,i}\right\|^{2}\leq P_{i}, \forall i\in\mathcal{K}_{\mathrm{R}},\label{DelayAware10e}
\end{align}
\end{subequations}
where the variables are $\left\{\mathbf{g}_{k,i},\mathrm{r}_{k,i}^{\mathrm{c}}\right\}_{k\in\mathcal{K}_{i}, i\in\mathcal{K}_{\mathrm{R}}}$ and $\left\{\epsilon_{k,i}\right\}_{k\in\widetilde{\mathcal{K}}_{i}, i\in\mathcal{K}_{\mathrm{R}}}$. The set $\widetilde{\mathcal{K}}_{i}$ is given by
$$\widetilde{\mathcal{K}}_{i}=\left\{k': k'\in\mathcal{K}_{\mathrm{U}}\setminus\mathcal{K}_{i}~\text{and}~c_{f_{k'},j}=1, \forall j\in\mathcal{K}_{\mathrm{R}}\setminus\left\{i\right\}\right\}.$$
and $\widehat{\mathrm{R}}_{k,i}^{\mathrm{c}}=\log_{2}\left(1+\widehat{\gamma}_{k,i}^{\mathrm{c}}\right)$ with $\widehat{\gamma}_{k,i}^{\mathrm{c}}$ redefined as
\begin{equation}\label{DelayAware11}
\widehat{\gamma}_{k,i}^{\mathrm{c}}=\frac{\left|c_{f_{k},i}\mathbf{h}_{k,i}^{H}\mathbf{g}_{k,i}\right|^{2}}
{\sum\limits_{k'\in\mathcal{K}_{i}\setminus\left\{k\right\}}
	\left|c_{f_{k'},i}\mathbf{h}_{k,i}^{H}\mathbf{g}_{k',i}\right|^{2}+\overline{\sigma}_{k,i}^{2}}
\end{equation}
where $\overline{\sigma}_{k,i}^{2}=\sum\limits_{j\in\mathcal{K}_{\mathrm{R}}\setminus\left\{i\right\}}\epsilon_{k,j}+\sigma^{2}$.  Constraint~\eqref{DelayAware10c} is activated at the optimal point of problem~\eqref{DelayAware10}. Unfortunately, problem~\eqref{DelayAware10} is still a nonconvex due to the existing of inter-user interference in the denominator of $\widehat{\gamma}_{k,i}^{\mathrm{c}}$ defined in~\eqref{DelayAware11}. To overcome this difficulty, we firstly reformulate problem~\eqref{DelayAware10} into the following equivalent form:
\begin{subequations}\label{DelayAware12}
\begin{align}
&\max~\sum\limits_{i\in\mathcal{K}_{\mathrm{R}}}\sum\limits_{k\in\mathcal{K}_{i}}\mathrm{r}_{k,i}^{\mathrm{c}},\label{DelayAware12a}\\
&s.t. ~\mathrm{r}_{k,i}^{\mathrm{c}}\leq\log_{2}\left(1+\overline{\gamma}_{k,i}^{\mathrm{c}}\right), \forall k\in\mathcal{K}_{i}, \forall i\in\mathcal{K}_{\mathrm{R}},\label{DelayAware12b}\\
&\overline{\gamma}_{k,i}^{\mathrm{c}}\leqslant\frac{\left|c_{f_{k},i}\mathbf{h}_{k,i}^{H}\mathbf{g}_{k,i}\right|^{2}}
{\chi_{k,i}}, \forall k\in\mathcal{K}_{i}, \forall i\in\mathcal{K}_{\mathrm{R}},\label{DelayAware12c}\\
&\sum\limits_{k\in\mathcal{K}_{i}}\left|c_{f_{k},i}\mathbf{h}_{k',i}^{H}\mathbf{g}_{k,i}\right|^{2}\leqslant\epsilon_{k',i}, \forall k'\in\widetilde{\mathcal{K}}_{i}, \forall i\in\mathcal{K}_{\mathrm{R}},\label{DelayAware12d}\\
&\sum\limits_{k'\in\mathcal{K}_{i}\setminus\left\{k\right\}}
\left|c_{f_{k'},i}\mathbf{h}_{k,i}^{H}\mathbf{g}_{k',i}\right|^{2}+\overline{\sigma}_{k,i}^{2}\leq\chi_{k,i}, \forall k\in\mathcal{K}_{i}, \forall i\in\mathcal{K}_{\mathrm{R}},\label{DelayAware12e}\\
&\tau\mathrm{r}_{k,i}^{\mathrm{c}}\leq S, \forall k\in\mathcal{K}_{i}, \forall i\in\mathcal{K}_{\mathrm{R}},\label{DelayAware12f}\\
&\sum\limits_{k\in\mathcal{K}_{i}}\left\|c_{f_{k},i}\mathbf{g}_{k,i}\right\|^{2}\leq P_{i}, \forall i\in\mathcal{K}_{\mathrm{R}},\label{DelayAware12g}
\end{align}
\end{subequations}
where the variables are $\left\{\mathbf{g}_{k,i},\mathrm{r}_{k,i}^{\mathrm{c}},\overline{\gamma}_{k,i}^{\mathrm{c}},\chi_{k,i}\right\}_{k\in\mathcal{K}_{i}, i\in\mathcal{K}_{\mathrm{R}}}$ and $\left\{\epsilon_{k,i}\right\}_{k\in\widetilde{\mathcal{K}}_{i}, i\in\mathcal{K}_{\mathrm{R}}}$. Note that in problem~\eqref{DelayAware12}, constraints~\eqref{DelayAware12c},~\eqref{DelayAware12d}, and~\eqref{DelayAware12e} are activated at the optimal point of problem~\eqref{DelayAware12}. Although the function $\frac{\left|c_{f_{k},i}\mathbf{h}_{k,i}^{H}\mathbf{g}_{k,i}\right|^{2}}{\chi_{k,i}}$ in~\eqref{DelayAware12c} is convex function, constraint~\eqref{DelayAware12c} is non-convex. Therefore, problem~\eqref{DelayAware12} is still non-convex. If $\widetilde{\mathcal{K}}_{i}=\emptyset$, the corresponding constraint given in~\eqref{DelayAware12d} can be removed, implying that the interference suppression among eRRH $i$, $\forall i\in\mathcal{K}_{\mathrm{R}}$ is not needed.

In what follows, we try to iteratively convexify the non-convex constraint~\eqref{DelayAware12c} such that an iterative optimization method is designed to address problem~\eqref{DelayAware12}. In light of~\cite{CLNguyen2015,OperalMarks1977,Bibby1974,JBeck2010}, we replace the right side of the inequality in~\eqref{DelayAware12c} by its lower bound, which now can be obtained by the first order approximation due to the convexity of $\frac{\left|c_{f_{k},i}\mathbf{h}_{k,i}^{H}\mathbf{g}_{k,i}\right|^{2}}{\chi_{k,i}}$. In particular, we have
\begin{equation}\label{DelayAware13}
\frac{\left|c_{f_{k},i}\mathbf{h}_{k,i}^{H}\mathbf{g}_{k,i}\right|^{2}}{\chi_{k,i}}\geq\Phi_{k,i}^{\left(I\right)}\left(\mathbf{g}_{k,i},\chi_{k,i}\right)
, \forall k\in\mathcal{K}_{i}, \forall i\in\mathcal{K}_{\mathrm{R}},
\end{equation}
where $\Phi_{k,i}^{\left(I\right)}\left(\mathbf{g}_{k,i},\chi_{k,i}\right)$ is defined as
\begin{equation}\label{DelayAware130}
\begin{split}
\Phi_{k,i}^{\left(I\right)}\left(\mathbf{g}_{k,i},\chi_{k,i}\right)
\triangleq&\frac{2\Re\left(c_{f_{k},i}\left(\mathbf{g}_{k,i}^{\left(I\right)}\right)^{\mathrm{H}}\mathbf{h}_{k,i}\mathbf{h}_{k,i}^\mathrm{H}
	\mathbf{g}_{k,i}\right)}{\chi_{k,i}^{\left(I\right)}}\\
&-\left(\frac{\left|c_{f_{k},i}\mathbf{h}_{k,i}^{H}\mathbf{g}_{k,i}^{\left(I\right)}\right|}
{\chi_{k,i}^{\left(I\right)}}\right)^{2}\chi_{k,i}.
\end{split}
\end{equation}
In~\eqref{DelayAware130}, $\mathbf{g}_{k,i}^{\left(I\right)}$ and $\chi^{\left(I\right)}$ denote the values of variables  $\mathbf{g}_{k,i}$ and $\chi$ obtained at the $I$-th iteration, respectively. Thus, we resort to solving iteratively problem~\eqref{DelayAware14} to obtain the solution to problem~\eqref{DelayAware10} instead of solving problem~\eqref{DelayAware12} directly.
\begin{subequations}\label{DelayAware14}
\begin{align}
&\max~\sum\limits_{i\in\mathcal{K}_{\mathrm{R}}}\sum\limits_{k\in\mathcal{K}_{i}}\mathrm{r}_{k,i}^{\mathrm{c}},\label{DelayAware14a}\\
s.t. ~&\eqref{DelayAware12b},\eqref{DelayAware12d},\eqref{DelayAware12e},\eqref{DelayAware12f},\eqref{DelayAware12g},\label{DelayAware14b}\\
&\overline{\gamma}_{k,i}^{\mathrm{c}}\leqslant\Phi_{k,i}^{\left(I\right)}\left(\mathbf{g}_{k,i},\chi_{k,i}\right), \forall k\in\mathcal{K}_{i}, \forall i\in\mathcal{K}_{\mathrm{R}},\label{DelayAware14c}
\end{align}
\end{subequations}
where the variables are $\left\{\mathbf{g}_{k,i},\mathrm{r}_{k,i}^{\mathrm{c}}, \overline{\gamma}_{k,i}^{\mathrm{c}}, \chi_{k,i}\right\}_{k\in\mathcal{K}_{i}, i\in\mathcal{K}_{\mathrm{R}}}$ and $\left\{\epsilon_{k,i}\right\}_{k\in\widetilde{\mathcal{K}}_{i}, i\in\mathcal{K}_{\mathrm{R}}}$. Problem~\eqref{DelayAware14} is a convex problem that can be easily solved with classical convex optimization solver, e.g., CVX~\cite{{AvialGrant2015}}. The detailed steps used to solve problem~\eqref{DelayAware10} are summarized as Algorithm~\ref{DelayAwareAlg01} where $\eta$ is a predefined stop threshold.
\begin{algorithm}[t]
\caption{Beamformer optimization for cache-level transmission}\label{DelayAwareAlg01}
\begin{algorithmic}[1]
\STATE Set $I=0$. Initialize non-zero $\mathbf{g}_{k,i}^{\left(I\right)}$ such that the power constraint is satisfied and  $\mathrm{r}_{k,i}^{c\left(I\right)}=0, \forall k\in\mathcal{K}_{i}, i\in\mathcal{K}_{\mathrm{R}}$. Compute $\overline{\gamma}_{k,i}^{c^{\left(I\right)}}$, $\chi_{k,i}^{\left(I\right)}$, and $\epsilon_{k',i}^{\left(I\right)}$ with $\mathbf{g}_{k,i}^{\left(I\right)}$, $k\in\mathcal{K}_{i}$, $k'\in\widetilde{\mathcal{K}}_{i}$, $i\in\mathcal{K}_{\mathrm{R}}$.\label{CachenableAlg0101}
\STATE  Solve~\eqref{DelayAware14} obtaining $\mathbf{g}_{k,i}^{\left(I+1\right)}$, $\mathrm{r}_{k,i}^{c\left(I+1\right)}$, $\overline{\gamma}_{k,i}^{c\left(I+1\right)}$,$\chi_{k,i}^{\left(I+1\right)}$, and $\epsilon_{k',i}^{\left(I+1\right)}$, $k\in\mathcal{K}_{i}$, $k'\in\widetilde{\mathcal{K}}_{i}$, $i\in\mathcal{K}_{\mathrm{R}}$ for the given $\mathbf{g}_{k,i}^{\left(I\right)}$ and $\chi_{k,i}^{\left(I\right)}$.\label{CachenableAlg0102}
\STATE If $\left| \sum\limits_{i\in\mathcal{K}_{\mathrm{R}}}\sum\limits_{k\in\mathcal{K}_{i}}\mathrm{r}_{k,i}^{c\left(I+1\right)}
-\sum\limits_{i\in\mathcal{K}_{\mathrm{R}}}\sum\limits_{k\in\mathcal{K}_{i}}\mathrm{r}_{k,i}^{c\left(I\right)}\right|\leqslant\eta$, stop iteration. Otherwise, set $I\leftarrow I+1$ and go to step~\ref{CachenableAlg0102}.\label{CachenableAlg0103}
\end{algorithmic}
\end{algorithm}

\setcounter{equation}{16}
\begin{figure*}
\hrulefill
\begin{equation}\label{DelayAware15}
\widetilde{\gamma}_{k,i}^{\mathrm{c}}=\frac{\left|c_{f_{k},i}\mathbf{h}_{k,i}^{H}\mathbf{g}_{k,i}\right|^{2}}
{\sum\limits_{k'\in\mathcal{K}_{i}\setminus\left\{k\right\}}\left|c_{f_{k'},i}\mathbf{h}_{k,i}^{H}\mathbf{g}_{k',i}\right|^{2}+
\sum\limits_{k'\in\widetilde{\mathcal{K}}_{i}}\sum\limits_{k\in\mathcal{K}_{i}}\left|c_{f_{k},i}
\mathbf{h}_{k',i}^{H}\mathbf{g}_{k,i}\right|^{2}+\sigma^{2}}.
\end{equation}
\hrulefill
\end{figure*}
Let $r^{\mathrm{c}}$ denote the objective value of problem~\eqref{DelayAware14}. If we replace $\mathbf{g}_{k,i}$, $\mathrm{r}_{k,i}^{\mathrm{c}}$, $\overline{\gamma}_{k,i}^{\mathrm{c}}$, $\chi_{k,i}$, and $\epsilon_{k',i}$ with $\mathbf{g}_{k,i}^{\left(I\right)}$, $\left(\mathrm{r}_{k,i}^{c}\right)^{\left(I\right)}$, $\left(\overline{\gamma}_{k,i}^{c}\right)^{\left(I\right)}$, $\chi_{k,i}^{\left(I\right)}$, and $\epsilon_{k',i}^{\left(I\right)}$, $k'\in\widetilde{\mathcal{K}}_{i}$, $k\in\mathcal{K}_{i}$, $i\in\mathcal{K}_{\mathrm{R}}$,  respectively, all constraints are still satisfied, which means that the optimal solution of the $I$-th iteration is feasible point of problem~\eqref{DelayAware14} in the $\left(I+1\right)$-th iteration. This is because of the approximation in~\eqref{DelayAware14c}~\cite{OperalMarks1977}. Thus, the objective obtained in the $\left(I+1\right)$-th iteration is no smaller than that in the $I$-th iteration, i.e., $\left(r^{\mathrm{c}}\right)^{\left(I+1\right)}\geq \left(r^{\mathrm{c}}\right)^{\left(I\right)}$. In other words, Algorithm~\ref{DelayAwareAlg01} generates a nondecreasing sequence of objective values $\left(r^{\mathrm{c}}\right)^{\left(I\right)}$. Moreover, the problem has an upper bound due to the power constraints. Therefore, the convergence of Algorithm~\ref{DelayAwareAlg01} can be guaranteed by the monotonic boundary theorem~\cite{Bibby1974}. Furthermore, following the same arguments as those in~\cite[Theorem 1]{OperalMarks1977}, we can prove that Algorithm~\ref{DelayAwareAlg01} converges to a Karush-Kuhn-Tucker (KKT) solution of problem~\eqref{DelayAware10}.

In Algorithm~\ref{DelayAwareAlg01}, Step~\ref{CachenableAlg0102} solves a convex problem, which can be efficiently implemented by primal-dual interior point method with approximate complexity of $\mathit{O}\left(\left(\overline{K}_{\mathrm{U}}\left(N_{\mathrm{tm}}+4\right)\right)^{3.5}\right)$, where $N_{\mathrm{tm}}=\max\limits_{i\in\mathcal{K}_{\mathrm{R}}}N_{\mathrm{t},i}$, $\overline{K}_{\mathrm{U}}=\sum\limits_{i\in\mathcal{K}_{\mathrm{R}}}\left|\mathcal{K}_{i}\right|$, and $\mathit{O}\left(\cdot\right)$ stands for the big-O notation~\cite{MathPotra2000}. The overall computational complexity is $\mathit{O}\left(\kappa_{1}\left(K_{\mathrm{U}}\left(N_{\mathrm{tm}}+4\right)\right)^{3.5}\right)$, where $\kappa_{1}$ denotes the number of the iteration of Algorithm~\ref{DelayAwareAlg01}.

\subsection*{B. Decentralized Optimization for Cache-level Transmission}
Though problem~\eqref{DelayAware08} can be solved by Algorithm~\ref{DelayAwareAlg01}, it needs to be implemented centrally at BBU, which results in large burden on fronthaul links. Note that the existing decentralized optimization methods that are designed to address the optimization problem for coordinated multiple point network need signaling exchange among BSs, which results in burden on backhaul link and incurs a certain delivery latency~\cite{TSPShi2011,TSPWeeraddana2013,TSPHuang2011,TSPTervo2018,TVTHuang2014}. From the perspective of reducing the delivery latency caused by information exchange and making the best use of the available signal processing capability of each eRRH, Algorithm~\ref{DelayAwareAlg01} and the existing decentralized optimization algorithms~\cite{TSPShi2011,TSPWeeraddana2013,TSPHuang2011,TSPTervo2018,TVTHuang2014}  are not suitable for cache-enabled F-RANs.

For F-RANs, the objective is to reduce delivery latency and the overhead of fronthaul links. With the help of the cached contents, each eRRH can design its transmitted signal based only on its local cached content.  In addition, the signalling exchange between eRRHs results in a large potential latency and requires stringent synchronization between eRRHs. Therefore, the optimization of transmit beamformers is expected to be independently implemented at each eRRH for cache-level transmission. The denominator of the right item in~\eqref{DelayAware04} implies that eRRH $i$ cannot independently optimize $\mathbf{g}_{k,i}$, $k\in\mathcal{K}_{i}$, due to the existing of inter-eRRH interference. In order to release the coupling among eRRHs and to achieve autonomous optimization of the beamformers, we introduce a new concept that is defined as the signal-to-interference-leakage-plus-noise ratio (SILNR), i.e.,~\eqref{DelayAware15}, at the top of the next page.
In the denominator of~\eqref{DelayAware15}, the first term denotes the intra-eRRH interference and the second term denotes the total power leaked from user $k\in\mathcal{K}_{i}$ to all other user $k'\in\widetilde{\mathcal{K}}_{i}$. It is not difficult to see that the SILNR $\widetilde{\gamma}_{k,i}^{\mathrm{c}}$ only depends on the local CSI, i.e., the channel coefficient from eRRH $i$ to all users, which can be easily obtained, $\forall i\in\mathcal{K}_{\mathrm{R}}$. To design an autonomous optimization, we replace the achievable rate $\mathrm{R}_{k,i}^{\mathrm{c}}$ in constraints~\eqref{DelayAware08b} with an approximated achievable rate $\widetilde{\mathrm{R}}_{k,i}^{\mathrm{c}}=\log_{2}\left(1+\widetilde{\gamma}_{k,i}^{\mathrm{c}}\right)$,  $\forall k\in\mathcal{K}_{i}$, $\forall i\in\mathcal{K}_{\mathrm{R}}$. Thus, instead of directly addressing problem~\eqref{DelayAware08}, the problem of beamformer optimization for the cache-level transmission is formulated as a series of parallel decentralized optimization problems given by~\eqref{DelayAware16} at eRRH $i$.
\setcounter{equation}{17}
\begin{subequations}\label{DelayAware16}
\begin{align}
&\max~\tau\sum\limits_{k\in\mathcal{K}_{i}}\mathrm{r}_{k,i}^{\mathrm{c}},\label{DelayAware16a}\\
s.t. ~&\mathrm{r}_{k,i}^{\mathrm{c}}\leq\widetilde{\mathrm{R}}_{k,i}^{\mathrm{c}}, \forall k\in\mathcal{K}_{i},\label{DelayAware16b}\\
&\tau\mathrm{r}_{k,i}^{\mathrm{c}}\leq S, \forall k\in\mathcal{K}_{i},\label{DelayAware16c}\\
&\sum\limits_{k\in\mathcal{K}_{i}}\left\|c_{f_{k},i}\mathbf{g}_{k,i}\right\|^{2}\leq P_{i},\label{DelayAware16d}
\end{align}
\end{subequations}
where the variable is $\left\{\mathbf{g}_{k,i},\mathrm{r}_{k,i}^{\mathrm{c}}\right\}_{k\in\mathcal{K}_{i}}$. The goal of problem~\eqref{DelayAware16} is to maximize the local delivery rate, while taking the inter-eRRH interference suppression into account, i.e., maximizing the total delivery rate of cache-level transmission as much as possible. Similarly, problem~\eqref{DelayAware16} can be equivalently transformed into the following form:
\begin{subequations}\label{DelayAware17}
\begin{align}
&\max~\sum\limits_{k\in\mathcal{K}_{i}}\mathrm{r}_{k,i}^{\mathrm{c}},\label{DelayAware17a}\\
s.t. ~&\mathrm{r}_{k,i}^{\mathrm{c}}\leq\log_{2}\left(1+\overline{\gamma}_{k,i}^{\mathrm{c}}\right), \forall k\in\mathcal{K}_{i},\label{DelayAware17b}\\
&\overline{\gamma}_{k,i}^{\mathrm{c}}\leqslant\frac{\left|c_{f_{k},i}\mathbf{h}_{k,i}^{H}\mathbf{g}_{k,i}\right|^{2}}
{\chi_{k,i}}, \forall k\in\mathcal{K}_{i},\label{DelayAware17c}\\
&\sum\limits_{k\in\mathcal{K}_{i}}\left\|c_{f_{k},i}\mathbf{h}_{k',i}^{H}\mathbf{g}_{k,i}\right\|^{2}\leqslant\epsilon_{k',i}, \forall k'\in\widetilde{\mathcal{K}}_{i},\label{DelayAware17d}\\
&\sum\limits_{k'\in\mathcal{K}_{i}\setminus\left\{k\right\}}
\left|c_{f_{k'},i}\mathbf{h}_{k,i}^{H}\mathbf{g}_{k',i}\right|^{2}+\widetilde{\sigma}_{k,i}^{2}\leq\chi_{k,i}, \forall k\in\mathcal{K}_{i},\label{DelayAware17e}\\
&\tau\mathrm{r}_{k,i}^{\mathrm{c}}\leq S, \forall k\in\mathcal{K}_{i}, \label{DelayAware17f}\\
&\sum\limits_{k\in\mathcal{K}_{i}}\left\|c_{f_{k},i}\mathbf{g}_{k,i}\right\|^{2}\leq P_{i},\label{DelayAware17g}
\end{align}
\end{subequations}
where the variables are $\left\{\mathbf{g}_{k,i},\mathrm{r}_{k,i}^{\mathrm{c}}, \overline{\gamma}_{k,i}^{\mathrm{c}}, \chi_{k,i}\right\}_{k\in\mathcal{K}_{i}}$ and $\left\{\epsilon_{k',i}\right\}_{k'\in\widetilde{\mathcal{K}}_{i}}$. $\widetilde{\sigma}_{k,i}^{2}=\sum\limits_{k'\in\widetilde{\mathcal{K}}_{i}}\epsilon_{k',i}+\sigma^{2}$.
Furthermore, problem~\eqref{DelayAware17} can be independently solved at each eRRH with similar procedure as Algorithm~\ref{DelayAwareAlg01} without any signaling exchange among eRRHs. The overall computational complexity is $\mathit{O}\left(\kappa_{i,2}\left(\left|\mathcal{K}_{i}\right|\left(N_{\mathrm{tm}}+3\right)+K_{\mathrm{U}}\right)^{3.5}\right)$, where $\kappa_{i,2}$ denotes the number of the iterations of decentralized optimization algorithm of problem~\eqref{DelayAware17}.

\begin{remark}
\rm Note that to obtain the performance gain, the centralized optimization algorithm needs to collect all channel state information (CSI) and data file to the central processing unit. However, the decentralized optimization algorithm needs only partial CSI that can be obtained by the training sequence of the uplink and optimize the beamforming vector without sharing data files. In time division duplex (TDD) system, each BS can directly estimate the CSI of its own users without any information exchange among eRRHs, by exploiting uplink and downlink reciprocity. In addition, each eRRH can estimate the crosstalk channels to other users served by other eRRHs from the reverse link.
\end{remark}
\section*{\sc \uppercase\expandafter{\romannumeral5}. Design of Optimization Algorithm for Network-level Transmission}

In this section,  we focus on addressing problem~\eqref{DelayAware09} with centralized way at the BBU. Different from the signal processing at each eRRHs,  due to the fact that the BBU has the whole CSI and data files of all users, the BBU can centrally design the precoders for all users and jointly transmit data files to each user. Thus, the intra- and inter-eRRH interference can be efficiently controlled to maximize the sum of delivery data rate. Following a similar procedure used to solve problem~\eqref{DelayAware10}, we firstly introduce some auxiliary variables $\iota_{k}$, $\mu_{k}$, $\forall k\in\mathcal{K}_{\mathrm{U}}$, and then reformulate equivalently problem~\eqref{DelayAware09} into the following form:
\begin{subequations}\label{DelayAware18}
\begin{align}
&\max~\sum\limits_{k\in\mathcal{K}_{\mathrm{U}}}\mathrm{r}_{k}^{\mathrm{n}},\label{DelayAware18a}\\
s.t. ~&\mathrm{r}_{k}^{\mathrm{n}}\leq\log_{2}\left(1+\iota_{k}\right),  \forall k\in\mathcal{K}_{U},\label{DelayAware18b}\\
&\iota_{k}\leq\frac{\left|\mathbf{h}_{k}^{H}\mathbf{u}_{k}\right|^{2}}{\mu_{k}},  \forall k\in\mathcal{K}_{U},\label{DelayAware18c}\\
&\sum\limits_{k'\in\mathcal{K}_{\mathrm{U}}, k'\neq k}\left|\mathbf{h}_{k}^{H}\mathbf{u}_{k'}\right|^{2}+\mathbf{h}_{k}^{H}\mathbf{\Omega}\mathbf{h}_{k}^{H}+\sigma^{2}\leq\mu_{k},  \forall k\in\mathcal{K}_{U},\label{DelayAware18d}\\
&\sum\limits_{k\in\mathcal{K}_{\mathrm{U}}}\left\|\mathbf{B}_{i}\mathbf{u}_{k}\right\|^{2}+\mathrm{tr}\left(\mathbf{\Omega}_{i}\right)\leq P_{i}, \forall i\in\mathcal{K}_{R},\label{DelayAware18e}\\
&\mathrm{r}_{k}^{\mathrm{n}}\leq \max\left(S-\tau\overline{\mathrm{r}}_{k}^{\mathrm{c}},0\right), \forall k\in\mathcal{K}_{U},\label{DelayAware18f}\\
&\log_{2}\left(\left|\overline{\mathbf{A}}_{i}\right|\right)-\log_{2}\left(\left|\mathbf{\Omega}_{i}\right|\right)\leqslant C_{i}, \forall i\in\mathcal{K}_{R},\label{DelayAware18g}
\end{align}
\end{subequations}
where the variable is $\left\{\mathbf{u}_{k},\mathrm{r}_{k}^{\mathrm{n}},\iota_{k},\mu_{k}, \mathbf{\Omega}_{i}\right\}_{k\in\mathcal{K}_{\mathrm{U}}, i\in\mathcal{K}_{\mathrm{R}}}$. Note that the constraints~\eqref{DelayAware18c} and~\eqref{DelayAware18d} are activated at the optimal point of problem~\eqref{DelayAware18}. The difficulties of solving problem~\eqref{DelayAware18} lie in~\eqref{DelayAware18c} and~\eqref{DelayAware18g}, as those contraints are non-convex. To overcome the non-convexity of the right hands of the inequalities in~\eqref{DelayAware18c},  we resort to the following inequality~\cite{CLNguyen2015}
\begin{equation}\label{DelayAware19}
\frac{\left|\mathbf{h}_{k}^{H}\mathbf{u}_{k}\right|^{2}}{\mu_{k}}\geq\psi_{k}^{\left(I\right)}\left(\mathbf{u}_{k},\mu_{k}\right), \forall k\in\mathcal{K}_{U},
\end{equation}
where $\psi_{k}^{\left(I\right)}\left(\mathbf{u}_{k},\mu_{k}\right)$ is defined as
\begin{equation}\label{DelayAware190}
\psi_{k}^{\left(I\right)}\left(\mathbf{u}_{k},\mu_{k}\right)
\triangleq\frac{2\Re\left(\left(\mathbf{u}_{k}^{\left(I\right)}\right)^{\mathrm{H}}\mathbf{h}_{k}\mathbf{h}_{k}^\mathrm{H}
	\mathbf{u}_{k}\right)}{\mu_{k}^{\left(I\right)}}-\left(\frac{\left|\mathbf{h}_{k}^{H}\mathbf{u}_{k}^{\left(I\right)}\right|}
{\mu_{k}^{\left(I\right)}}\right)^{2}\mu_{k}.
\end{equation}
To convexify the left side of constraint~\eqref{DelayAware18g}, we need to transform the first function $\log_{2}\left(\cdot\right)$ into a convex form. According to the concavity property of function $\log_{2}\left(\cdot\right)$, we have
\begin{equation}\label{DelayAware20}
\log_{2}\left(\left|\overline{\mathbf{A}}_{i}\right|\right)\leq\varphi\left(\overline{\mathbf{A}}_{i},\overline{\mathbf{B}}_{i}\right)
\end{equation}
where
$\overline{\mathbf{B}}_{i}=\mathbf{B}_{i}\sum\limits_{k\in\mathcal{K}_{\mathrm{U}}}\mathbf{u}_{k}^{\left(I\right)}\left(\mathbf{u}_{k}^{\left(I\right)}\right)^{\mathrm{H}}\mathbf{B}_{i}^{\mathrm{H}}+\mathbf{\Omega}_{i}^{\left(I\right)}$ and
 $\varphi\left(\mathbf{A},\mathbf{B}\right)=\log_{2}\left(\left|\mathbf{B}\right|\right)+\frac{1}{\ln\left(2\right)}\mathrm{tr}\left(\mathbf{B}^{-1}\left(\mathbf{A}-\mathbf{B}\right)\right)$. Now, instead of directly solving problem~\eqref{DelayAware18}, we iteratively solve the following problem to obtain its solution by replacing the right hands of the inequalities in~\eqref{DelayAware18c} with $\psi_{k}^{\left(I\right)}\left(\mathbf{u}_{k},\mu_{k}\right)$, $\forall k\in\mathcal{K}_{U}$ and $\log_{2}\left(\left|\overline{\mathbf{A}}_{i}\right|\right)$ with the right side of~\eqref{DelayAware20} .
\begin{subequations}\label{DelayAware21}
\begin{align}
&\max~\sum\limits_{k\in\mathcal{K}_{\mathrm{U}}}\mathrm{r}_{k}^{\mathrm{n}},\label{DelayAware21a}\\
s.t. ~&\eqref{DelayAware18b},\eqref{DelayAware18d},\eqref{DelayAware18e},\eqref{DelayAware18f},\label{DelayAware21b}\\
&\iota_{k}\leq\psi_{k}^{\left(I\right)}\left(\mathbf{u}_{k},\mu_{k}\right), \forall k\in\mathcal{K}_{\mathrm{U}},\label{DelayAware21c}\\
&\varphi\left(\overline{\mathbf{A}}_{i},\overline{\mathbf{B}}_{i}\right)-\log_{2}\left(\left|\mathbf{\Omega}_{i}\right|\right)\leqslant C_{i}, \forall i\in\mathcal{K}_{\mathrm{R}},\label{DelayAware21d}
\end{align}
\end{subequations}
where the variable is $\left\{\mathbf{u}_{k},\mathrm{r}_{k}^{\mathrm{n}},\iota_{k},\mu_{k}, \mathbf{\Omega}_{i}\right\}_{k\in\mathcal{K}_{\mathrm{U}}, i\in\mathcal{K}_{\mathrm{R}}}$. After applying the convexity operations, problem~\eqref{DelayAware18} is transformed into approximated convex problem~\eqref{DelayAware21} that can be easily solved with convex optimization tools. A step-by-step summary of the beamforming design is given in Algorithm~\ref{DelayAwareAlg02} for network-level transmission. Note that when $\tau$ equals to zero, problem~\eqref{DelayAware09} is equivalent to the optimization of beamforming vectors in traditional cloud-based coordinated joint transmission, which means that Algorithm~\ref{DelayAwareAlg02} can also be used for the optimization of traditional cloud-based coordinated joint transmission systems.
\begin{algorithm}[t]
\caption{Beamformer optimization for network-level transmission}\label{DelayAwareAlg02}
\begin{algorithmic}[1]
\STATE Set $I=0$. Initialize non-zero $\mathbf{u}_{k}^{\left(I\right)}$ such that power constraint is satisfied and  $\mathrm{r}_{k}^{n\left(I\right)}=0$, $\forall k\in\mathcal{K}_{U}$. Compute $\iota_{k}^{\left(I\right)}$ and $\mu_{k}^{\left(I\right)}$ with $\mathbf{u}_{k}^{\left(I\right)}$, $k\in\mathcal{K}_{U}$.\label{CachenableAlg0201}
\STATE  Solve~\eqref{DelayAware21} obtaining $\mathbf{u}_{k}^{\left(I+1\right)}$, $\mathrm{r}_{k}^{n\left(I+1\right)}$, $\iota_{k}^{\left(I+1\right)}$, and $\mu_{k}^{\left(I+1\right)}$, $k\in\mathcal{K}_{U}$ for the given $\mathbf{u}_{k}^{\left(I\right)}$ and $\mu_{k}^{\left(I\right)}$.\label{CachenableAlg0202}
\STATE If $\left| \sum\limits_{k\in\mathcal{K}_{U}}\mathrm{r}_{k}^{n\left(I+1\right)}
-\sum\limits_{k\in\mathcal{K}_{i}}\mathrm{r}_{k}^{n\left(I\right)}\right|\leqslant\eta$, then stop iteration. Otherwise, set $I\leftarrow I+1$ and go to step~\ref{CachenableAlg0202}.\label{CachenableAlg0203}
\end{algorithmic}
\end{algorithm}

Similarly, the sequence generated by Algorithm~\ref{DelayAwareAlg02} is also nondecreasing. In other words, Algorithm~\ref{DelayAwareAlg02} also converges to a KKT solution of problem~\eqref{DelayAware09}. In Algorithm~\ref{DelayAwareAlg02}, Step~\ref{CachenableAlg0102} solves a convex optimization problem, which can be efficiently implemented by the primal-dual interior point method with approximate complexity of $\mathit{O}\left(\left(K_{\mathrm{U}}\left(N_{\mathrm{tt}}+3\right)+
K_{\mathrm{R}}N_{\mathrm{tm}}^{2}\right)^{3.5}\right)$~\cite{MathPotra2000}. Suppose Algorithm~\ref{DelayAwareAlg02} takes totally $\kappa_{3}$ number of the iteration to converge. The overall computational complexity is $\mathit{O}\left(\kappa_{3}\left(K_{\mathrm{U}}\left(N_{\mathrm{tt}}+3\right)+K_{\mathrm{R}}N_{\mathrm{tm}}^{2}\right)^{3.5}\right)$.
It can be observed that among our developed algorithms, the computational complexity of network-level optimization algorithm is higher than that of cache-level optimization algorithm and is similar with the proposed algorithms in~\cite{TWCPark2016}.

\section*{\sc \uppercase\expandafter{\romannumeral6}. Numerical Results}

In this section, we present some numerical results to evaluate the performance of proposed algorithms. We consider a cache-enabled F-RAN where the positions of eRRHs and UEs are uniformly distributed within a circular cell of radius of $500$ m. The channel vector $\mathbf{h}_{k,i}$ from eRRH $i$ to the user $k$ is modeled as $\mathbf{h}_{k,i}=\sqrt{\varrho_{k,i}}\widetilde{\mathbf{h}}_{k,i}$ , where the channel power $\varrho_{k,i}$ is given as $\varrho_{k,i} = 1/\left(1 + (d_{k,i}/d_{0}\right)^{\alpha})$ and the elements of $\widetilde{\mathbf{h}}_{k,i}$ are independent and identically distributed (i.i.d.) with zero mean and unit variance. For simplicity, we assume that the eRRHs have the same number of transmit antennas, the same transmit power, the same fronthaul capacity, and the same size of cache, i.e., $N_{\mathrm{t},i}=N_{\mathrm{t}}$, $P_{i}=P$, $C_{i}=C$ and $B_{i}=B$, $\forall i\in\mathcal{K}_{\mathrm{R}}$. If not stated elsewhere, the simulation is performed with the parameters given in Table~\ref{SimulatedParametersValues}.
\begin{table}[htbp]
\renewcommand{\captionfont}{\footnotesize}
\renewcommand*\captionlabeldelim{.}
	\setlength{\abovecaptionskip}{0pt}
	\setlength{\belowcaptionskip}{5pt}
	\captionstyle{flushleft}
	\onelinecaptionstrue
	\centering
	\caption{Simulation parameters}
	\begin{tabular}{|c|c||c|c|}
		\hline
		Symbol&\makecell[c]{Value}&Symbol&\makecell[c]{Value}\\
		\hline
		\hline
		$K_{\mathrm{R}}$& $3$&$K_{\mathrm{U}}$ &$5$\\
		\hline
		$N_{\mathrm{t}}$& $4$& $F$ & $10$\\
		\hline
		$B$ & $3S$&$\sigma^{2}$ &$1$\\
		\hline
		$d_{0}$ &$50$ m&$\alpha$ &$3$\\
		\hline
		$\eta$ &$10^{-3}$&$N_{s}$ &$2$\\
		\hline
	\end{tabular}
	\label{SimulatedParametersValues}
\end{table}

In the figures, ``CO-CEHTM'' and ``DO-CEHTM'' denote respectively the centralized optimization and decentralized optimization for cache-level transmission in our developed scheme. ``TC-CJTM'' denotes the traditional cloud-based coordinated joint transmission mechanism, i.e., $\tau=0$ in Algorithm~\ref{DelayAwareAlg02}. We assume that all files have the same degree of popularity. Namely, the cache state information $c_{f,i}$, $f\in\mathcal{F}$, $i\in\mathcal{K}_{\mathrm{R}}$, is randomly given subject to the cache capacity constraint and satisfies the conditions $\sum\limits_{i\in\mathcal{K}_{\mathrm{R}}}c_{f,i}=0$ or $\sum\limits_{i\in\mathcal{K}_{\mathrm{R}}}c_{f,i}=1$ while $\sum\limits_{f\in\mathcal{F}}c_{f,i}= \lfloor B/S\rfloor$. The requested file index $f_{k}$ of user $k$ is random selected from set $\mathcal{F}$. For cache-level transmission of full caching, the user set served by each eRRH does not intersect each other, i.e., each user is served by only one eRRH. All simulation results are obtained by averaging over $1000$ random channel realizations.

\begin{figure}[h]
\renewcommand{\captionfont}{\footnotesize}
\renewcommand*\captionlabeldelim{.}
\centering
\captionstyle{flushleft}
\onelinecaptionstrue
\includegraphics[width=1\columnwidth,keepaspectratio]{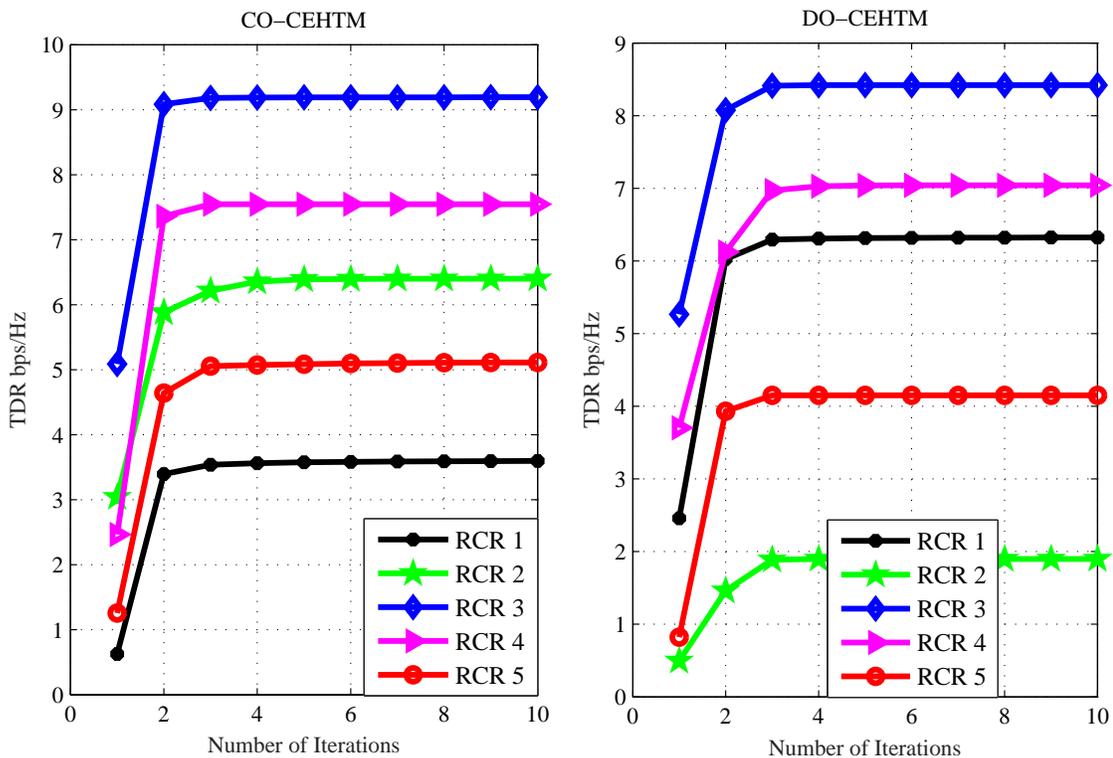}\\
\caption{Convergence trajectory of presented algorithms, $C=5$ bits/symbol, $S=10$, and $P=20$ dB.}
\label{ConvergenceTrajectory01}
\end{figure}

\begin{figure}[h]
\renewcommand{\captionfont}{\footnotesize}
\renewcommand*\captionlabeldelim{.}
\centering
\captionstyle{flushleft}
\onelinecaptionstrue
\includegraphics[width=1\columnwidth,keepaspectratio]{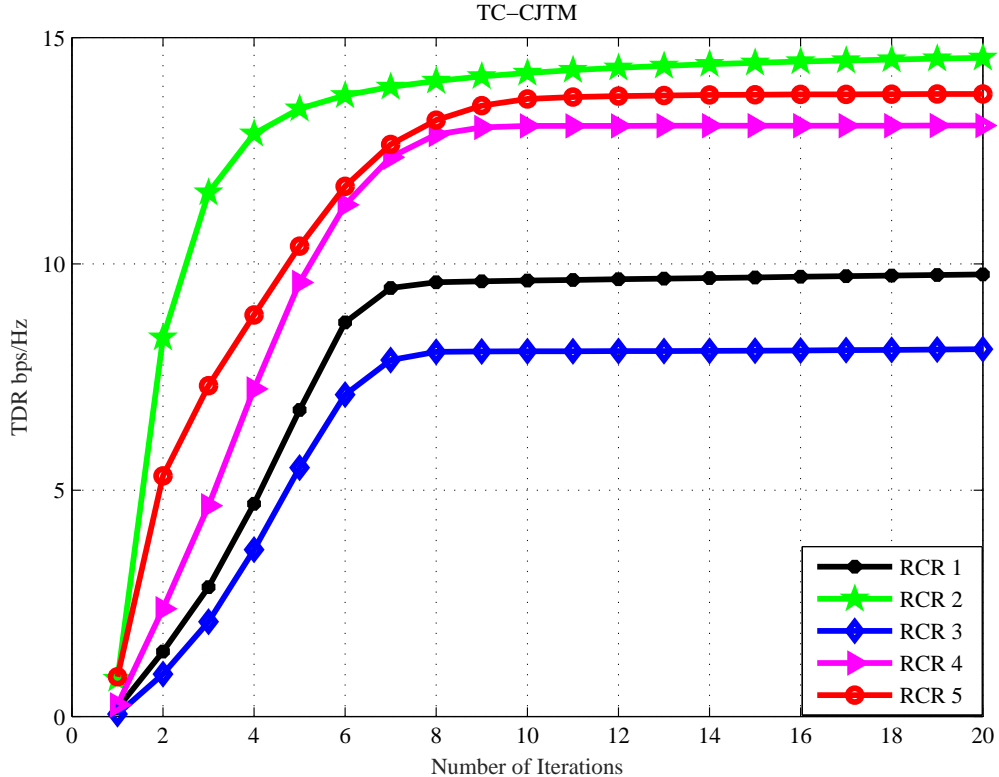}\\
\caption{Convergence trajectory of presented algorithms, $C=5$ bits/symbol, $S=10$, and $P=20$ dB.}
\label{ConvergenceTrajectory02}
\end{figure}

Fig.~\ref{ConvergenceTrajectory01} and Fig.~\ref{ConvergenceTrajectory02} illustrates respectively the convergence trajectory of the above mentioned three algorithms for a few randomly channel realizations (RCR) where $\tau=1$ ms for CO-CEHTM and DO-CEHTM and $\tau=0$ ms for TC-CJTM. It is seen that all algorithms generate a non-decreasing total delivery rate (TDR) sequence. Recalling that all the objective functions of problem~\eqref{DelayAware10},~\eqref{DelayAware18}, and~\eqref{DelayAware18} are boundary function due to the limitation of maximum allowable transmission power. The convergence of all algorithms can be guaranteed by the monotonic boundary theorem~\cite{Bibby1974}. Furthermore, it is observed that all algorithms can fast converge to a stationary point within limited number iterations, e.g., $10$ iterations.

\begin{figure}[h]
\renewcommand{\captionfont}{\footnotesize}
\renewcommand*\captionlabeldelim{.}
\centering
\captionstyle{flushleft}
\onelinecaptionstrue
\includegraphics[width=1\columnwidth,keepaspectratio]{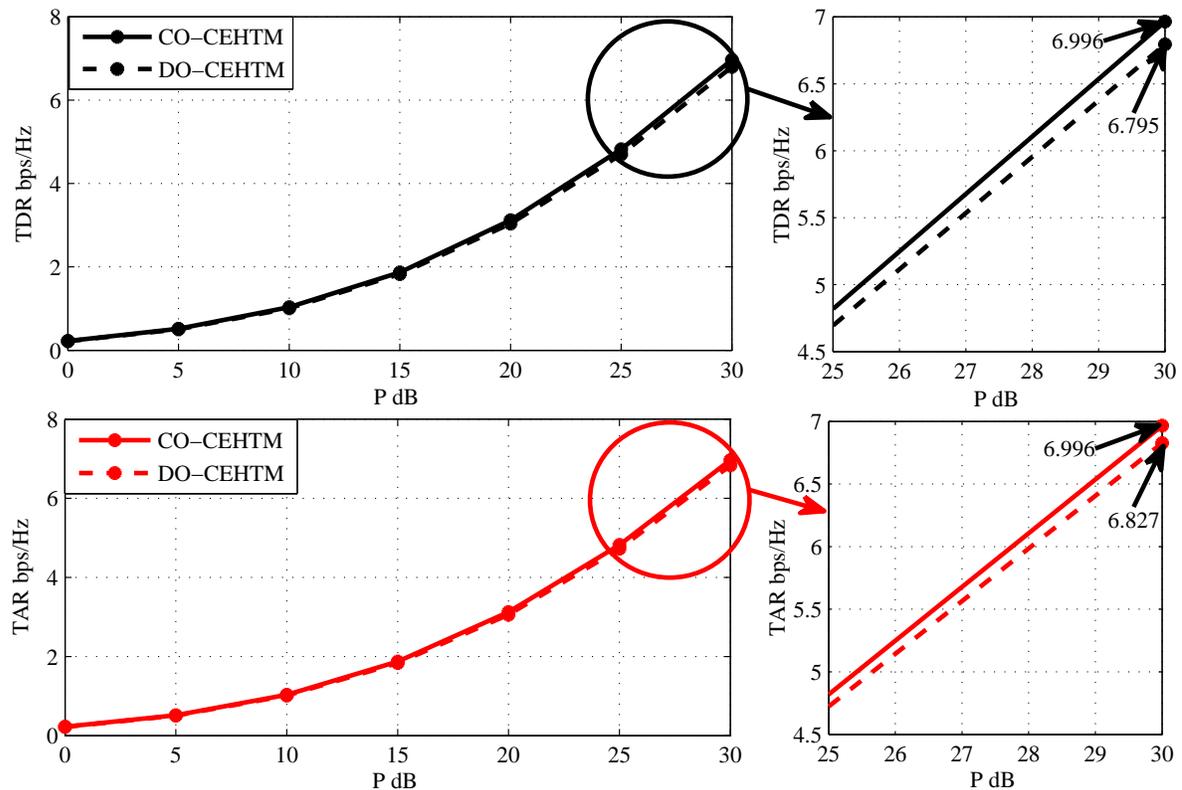}\\
\caption{Comparison of the average TDR and TAR of centralized and decentralized optimization algorithms for cache-level transmission, $\tau=0.01$ s, $C=5$ bit/symbol, and $S=10$.}
\label{ComparisonofCentravsDecentralized}
\end{figure}

Fig.~\ref{ComparisonofCentravsDecentralized} shows the comparison of average total delivery rate (TDR) and total achievable data rate (TAR) for centralized and decentralized optimization algorithms for cache-level transmission, respectively. It is shown that the performance of the decentralized and centralized algorithms is very close. The decentralized optimization algorithm can obtain more than $97\%$ of the performance obtained by the centralized optimization algorithm in terms of TDR and TAR, respectively. The performance gain of the centralized optimization algorithm is obtained at the cost of large number of backhaul overhead and CSI collection. In other words, the decentralized optimization algorithm is effective for delay sensitive data service and communication system with limited backhaul/fronthaul link.

\begin{figure}[h]
\renewcommand{\captionfont}{\footnotesize}
\renewcommand*\captionlabeldelim{.}
	\centering
	\captionstyle{flushleft}
	\onelinecaptionstrue
	\includegraphics[width=1\columnwidth,keepaspectratio]{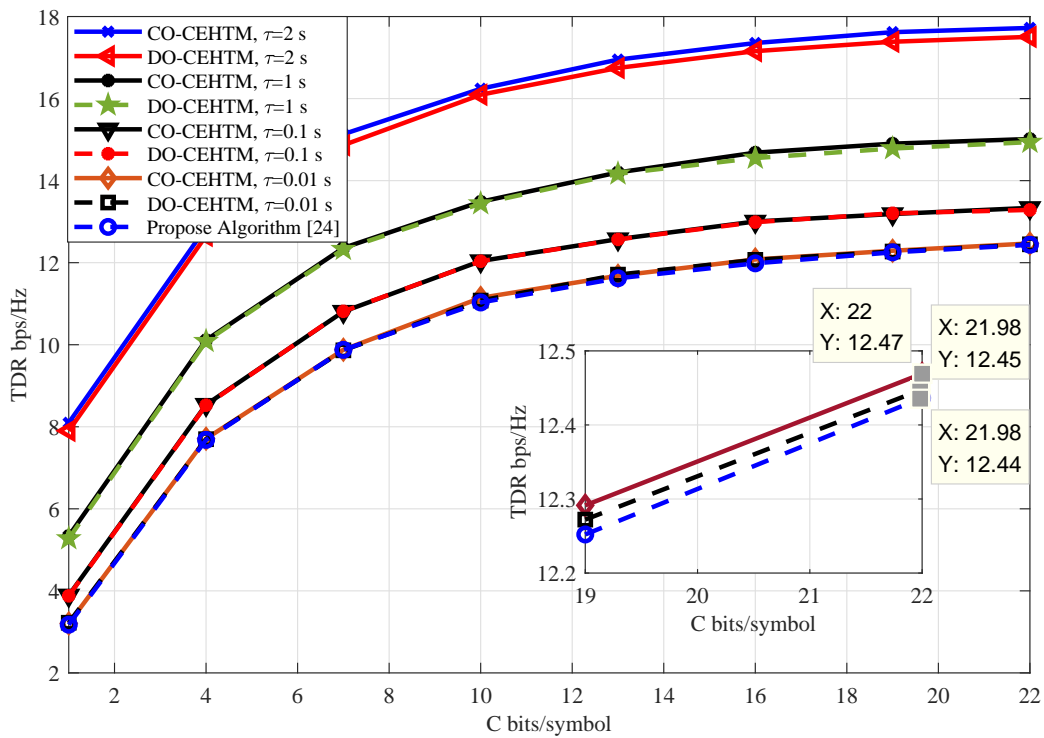}\\
	\caption{Comparison of the average TDR of hybrid transmission with centralized and decentralized optimization for cache-level transmission, $P=20$ dB and $S=10$.}
	\label{TDRvervusCapacity}
\end{figure}

Fig.~\ref{TDRvervusCapacity} illustrates the average TDR versus the fronthaul capacity $C$ for cache-enabled F-RANs with different delays. From this figure, we can observe that the average TDR increases with a larger capacity $C$ of fronthaul link in the region of small to middle fronthaul capacity, in which the performance is limited by the capacity of fronthaul link rather than the file size and the achievable data rate. However, for large capacity of fronthaul link, the average TDR performance is limited by other factors instead of the fronthaul capacity $C$ and becomes saturated. In addition, the centralized and decentralized optimization for cache-level transmission generate very close performance. Furthermore, increasing the delay results in increasing average TDR. The reason is that the duration of independent transmission at cache-level transmission or network level transmission increases with an increasing delay. In addition, the optimization of transmit beamforming vectors is transferred from eRRHs to the BBU such that the intra-cell interference and inter-cell interference are more effectively controlled. One can also note that the proposed two-level transmission scheme outperforms the algorithm proposed in~\cite{TWCPark2016} due to the delay $\tau$ which is used to transmit requested files.

\begin{figure}[h]
\renewcommand{\captionfont}{\footnotesize}
\renewcommand*\captionlabeldelim{.}
	\centering
	\captionstyle{flushleft}
	\onelinecaptionstrue
	\includegraphics[width=1\columnwidth,keepaspectratio]{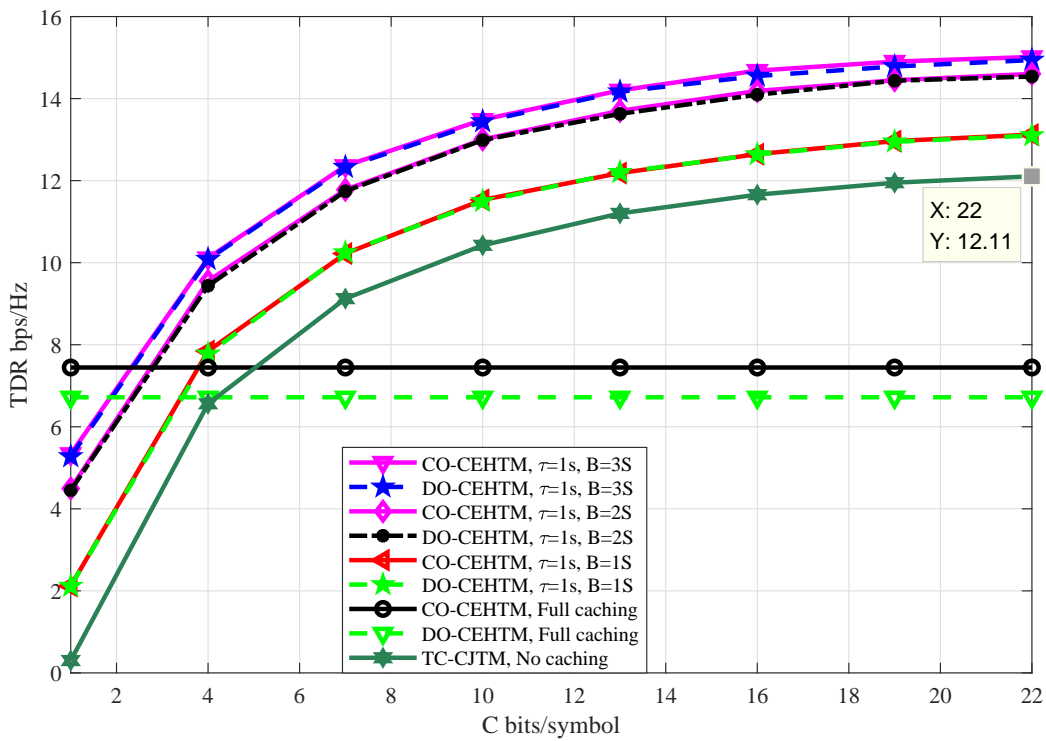}\\
	\caption{Comparison of the average TDR of hybrid transmission with centralized and decentralized optimization for cache-level transmission, $P=20$ dB and $S=10$.}
	\label{TDRvervusCapacity2}
\end{figure}

In Fig.~\ref{TDRvervusCapacity2}, we further evaluate the impact of the caching capacity $B$ on the average TDR versus the fronthaul capacity $C$ for cache-enabled F-RANs.  When the fronthaul capacity is sufficiently small, full caching provides a better way for cache-enabled F-RANs. We note that the proposed two-level transmission scheme outperforms the TC-CJTM due to the fact that the proposed two-level transmission scheme exploits fully the duration of the  delay $\tau$ to transmit data to partial users. Together with Fig.~\ref{TDRvervusCapacity}, it is shown that even if the delay $\tau$ is $10$ ms, the proposed two-level transmission scheme still obtain a certain performance gain compared to the TC-CJTM. In addition, one can see that different caching capacity $B$ have different impact on the TDR performance of the proposed two-level transmission scheme. It means that how to cache files and how many files to cache at each eRRH are also key problems for cache-enabled F-RANs and caching all files at eRRHs does not necessarily obtain the optimal system performance.

\begin{figure}[h]
\renewcommand{\captionfont}{\footnotesize}
\renewcommand*\captionlabeldelim{.}
	\centering
	\captionstyle{flushleft}
	\onelinecaptionstrue
	\includegraphics[width=1\columnwidth,keepaspectratio]{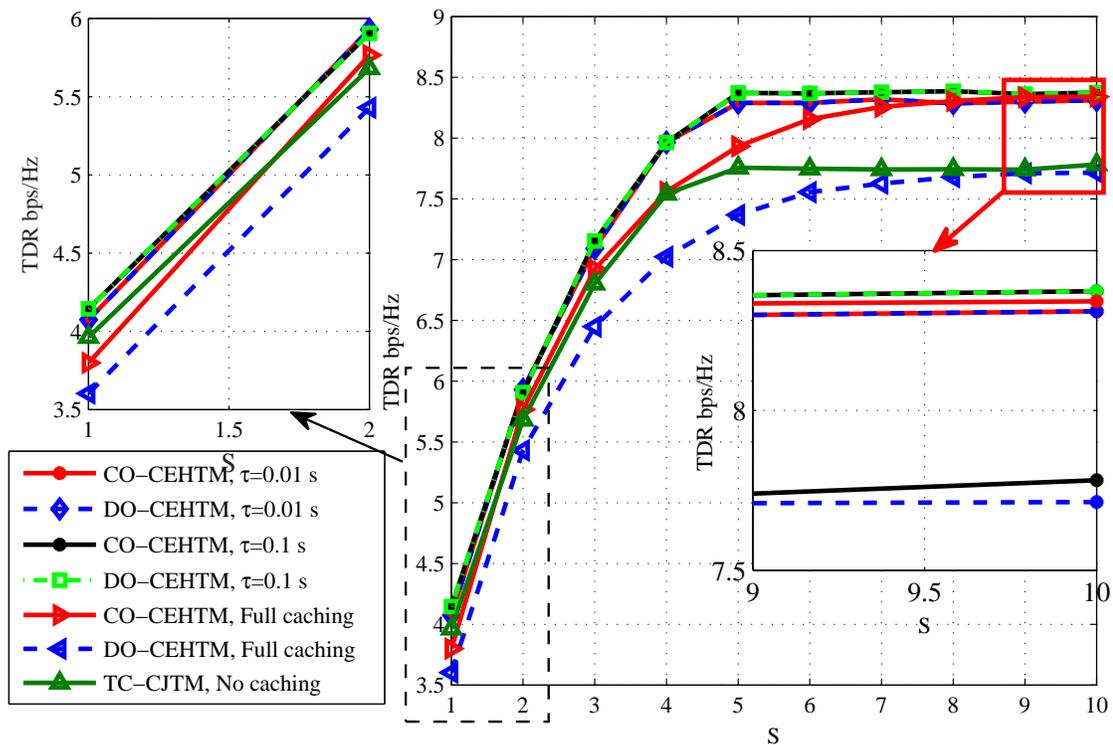}\\
	\caption{Comparison of the average TDR of hybrid transmission with centralized and decentralized optimization for cache-level transmission,  $P=20$ dB and $C=5$ bits/symbol.}
	\label{TDRvervusFileSizes}
\end{figure}

In Fig.~\ref{TDRvervusFileSizes}, we show the average TDR versus the file size $S$ for cache-enabled F-RANs. Similar to the observation in Fig.~\ref{TDRvervusCapacity}, the performance of TDR improves with a larger file size $S$ in the region of small fize sizes, in which the performance is limited by the file size $S$ rather than other factors, such as the fronthaul capacity $C$, the transmit power $P$, and the achievable data rates. In the region of medium to large file sizes, the TDR becomes saturated where the TDR performance is limited by the fronthaul capacity $C$, the transmit power $P$, or the achievable data rates. Note that for the full caching, compared to Fig.~\ref{ComparisonofCentravsDecentralized}, the performance gap between the centralized and decentralized optimization for the cache-level transmission increases due to the fact that the inter-eRRH interference suppression is limited to the number of the simultaneously served users. Numerical results further reveal that only when the file size is very small, the TDR performance of TC-CJTM is better than that of CO-CEHTM with full caching. It is due to the fact that the performance of TC-CJTM is constrained by the fronthaul capacity $C$, which does not affect the performance of CO-CEHTM. The proposed two-level transmission scheme with proper latency outperforms the transmission schemes with full or no caching.

\begin{figure}[h]
\renewcommand{\captionfont}{\footnotesize}
\renewcommand*\captionlabeldelim{.}
	\centering
	\captionstyle{flushleft}
	\onelinecaptionstrue
	\includegraphics[width=1\columnwidth,keepaspectratio]{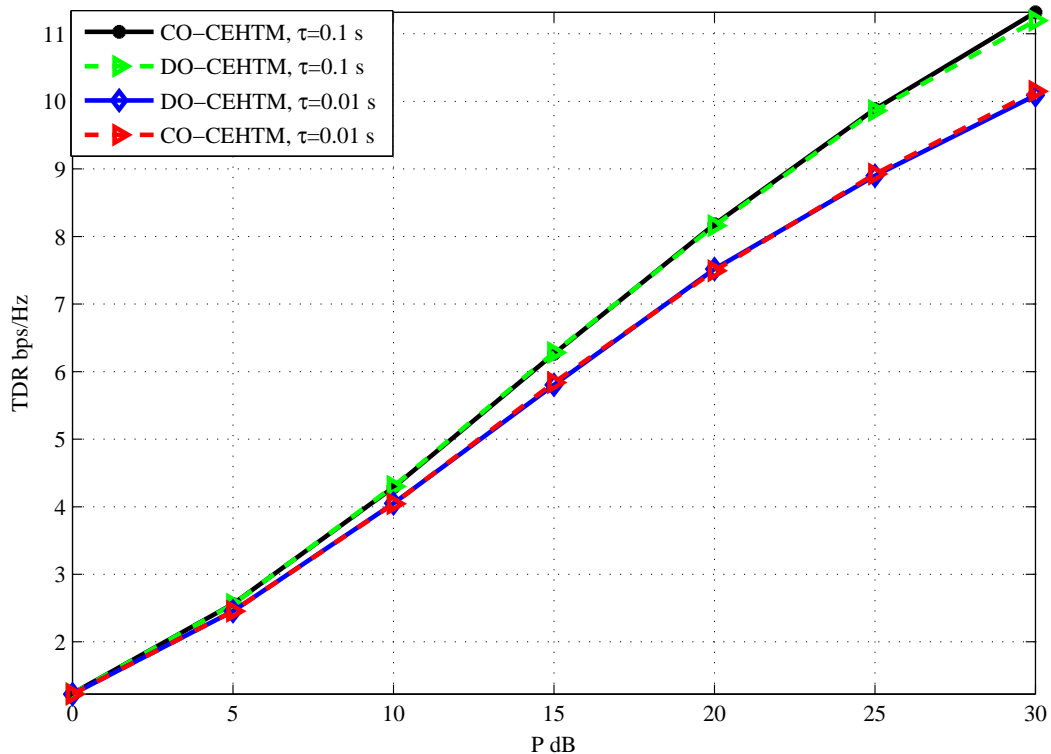}\\
	\caption{Comparison of the average TDR of hybrid transmission with centralized and decentralized optimization for cache-level transmission, $C=5$ bits/symbol, and $S=10$.}
	\label{TDRvervusPower}
\end{figure}

Fig.~\ref{TDRvervusPower} demonstrates the average performance of TDR versus maximum allowable transmit power $P$ for the downlink cache-enabled F-RANs. Numerical results also show that the performance of the centralized and decentralized optimization of cache-level transmission is very close. Furhtermore, the performance of the TDR increases with an increasing maximum transmit power $P$. In other words, in a certain power range, the TDR performance is not saturated.

\section*{\sc \uppercase\expandafter{\romannumeral6}. Conclusions}
In this paper, a two-level transmission scheme including cache-level and network-level transmission has been proposed for cache-enabled F-RANs under the constraints of the fronthaul capacity and the maximum allowable transmit power of eRRHs. Centralized optimization algorithms has been developed to address the problem of interest for cache-level and network-level transmission, respectively. To avoid the signaling exchange among eRRHs, a decentralized optimization algorithm has been further developed for cache-level transmission. Finally, numerical results have been provided to evaluate the performance of our developed algorithms.



\begin{small}

\end{small}

\end{document}